%% file: Toffoli-improved.tex
\documentclass[submission,copyright,creativecommons]{eptcs}
\usepackage{breakurl}
\usepackage{underscore}

\usepackage{amssymb,nicefrac}
\usepackage{amsthm}
\usepackage{amsfonts}
\usepackage{mathtools}
\usepackage{calc}
\usepackage{multicol}
\usepackage{hyperref}
\usepackage{graphicx}

\usepackage{amsmath}
\usepackage{stmaryrd}
\usepackage{xspace}
\usepackage{vwcol}
\usepackage{needspace}

\theoremstyle{plain}
\newtheorem{theorem}{Theorem}
\newtheorem{proposition}[theorem]{Proposition}
\newtheorem{corollary}[theorem]{Corollary}
\newtheorem{lemma}[theorem]{Lemma}

\theoremstyle{definition}
\newtheorem{definition}[theorem]{Definition}

\theoremstyle{remark}
\newtheorem*{remark}{Remark}

\usepackage{tikz}
\usepackage{pgfplots}
\usetikzlibrary{decorations.markings}
\usetikzlibrary{matrix,backgrounds,folding}
\usetikzlibrary{shapes.geometric}
\pgfdeclarelayer{edgelayer}
\pgfdeclarelayer{nodelayer}
\pgfsetlayers{background,edgelayer,nodelayer,main}
\tikzstyle{every picture}=[baseline=-0.25em]
\tikzstyle{none}=[inner sep=0mm]
\tikzstyle{zxnode}=[shape=circle, minimum width=.25cm, inner sep=0.5pt, font=\footnotesize, draw=black]
\tikzstyle{gn}=[zxnode ,fill=green]
\tikzstyle{rn}=[zxnode ,fill=red]
\tikzstyle{H box}=[rectangle,fill=yellow,draw=black,xscale=1,yscale=1,font=\small,inner sep=0.75pt,minimum width=0.15cm,minimum height=0.15cm]
\tikzstyle{ug}=[regular polygon, regular polygon sides=3, fill=red, draw=black,inner sep = 0pt,minimum width=1em]
\tikzstyle{black dot}=[inner sep=0.7mm,minimum width=0pt,minimum height=0pt,fill=black,draw=black,shape=circle]
\tikzstyle{dot}=[black dot]
\tikzstyle{white dot}=[dot,fill=white]
\tikzstyle{zwcross}=[diamond, draw, fill=gray, minimum width=0em, inner sep=1.5pt]

\tikzstyle{st}=[star,star points = 5, fill=white,draw=black,inner sep = 1.2pt,line width=1.2pt]
\tikzstyle{uglabel}=[rounded corners=0.2em,fill=green!20,inner sep=0.1em,font=\scriptsize, anchor=west, xshift=-0.2em, yshift=0,opacity=1]

\pgfdeclarelayer{edgelayer}
\pgfdeclarelayer{nodelayer}
\pgfsetlayers{background,edgelayer,nodelayer,main}
\tikzstyle{none}=[inner sep=0mm]
\tikzstyle{every loop}=[]

\def\fig{}

\newcounter{steps}%
\newcommand\step{\refstepcounter{steps}\thesteps}%
\def\thesteps{\ensuremath{\roman{steps})}}%

\newcommand{\eq}[2][~~]{
#1
\underset{\substack{#2}}{=}
#1
}

\newcommand{\equi}[2][\quad]{
#1
\underset{\substack{#2}}{\iff}
#1
}

\newcommand{\interp}[1] {\left\llbracket #1 \right\rrbracket}
\newcommand{\frag}[1]{$\frac{\pi}{#1}$-frag\-ment}
\newcommand{\titlerule}[1]{\begin{minipage}{\columnwidth}\begin{center}
\rule{(\textwidth-\widthof{#1})/2}{0.5pt}#1\rule{(\textwidth-\widthof{#1})/2}{0.5pt}
\end{center}
\end{minipage}}
\newcommand{\annoted}[3]{{\scriptstyle #1}\left\lbrace\mathrlap{\phantom{#3}}\right.\overbrace{#3}^{#2}}

\newcommand{\ket}[1]{\ensuremath{\left|  #1 \right\rangle}}
\def \zx {\textnormal{ZX}\xspace}
\def \dzx {\textnormal{$\mathrm{\Delta}$ZX}\xspace}
\def \zwh {\text{$\textnormal{ZW}_{\!\!\frac 1{\sqrt{2}}}$}\xspace}
\def \zw {\textnormal{ZW}\xspace}

\newcommand{\half}{\begin{tikzpicture}
		\node [style=st] (0) at (0,0) {};
\end{tikzpicture}}
\newcommand{\two}{\begin{tikzpicture}
		\draw (0,0) circle (0.25) ;
\end{tikzpicture}}
\newcommand{\zero}{\begin{tikzpicture}
		\node [style=dot] (0) at (0,0) {};
\end{tikzpicture}}

\newcommand{\callrule}[2]{\hyperlink{r:#1}{\textnormal{(#2)}}\xspace}

\newcommand{\so}{\callrule{rules}{S1}}
\newcommand{\st}{\callrule{rules}{S2}}
\newcommand{\bo}{\callrule{rules}{B1}}
\newcommand{\bt}{\callrule{rules}{B2}}
\newcommand{\hl}{\callrule{rules}{HL}}
\newcommand{\h}{\callrule{rules}{H}}
\newcommand{\iv}{\callrule{rules}{IV}}
\newcommand{\tz}{\callrule{rules}{T0}}
\newcommand{\bw}{\callrule{rules}{BW}}
\newcommand{\htt}{\callrule{rules}{HT}}
\newcommand{\tcx}{\callrule{rules}{TCX}}
\newcommand{\tw}{\callrule{rules}{TW}}
\newcommand{\z}{\callrule{rules}{Z}}
\newcommand{\e}{\callrule{clifford-t-toffoli}{E}}
\newcommand{\kt}{\callrule{clifford-t-toffoli}{K}}
\newcommand{\p}{\callrule{clifford-t-toffoli}{P}}

\newcommand{\scomp}{\frac{\pi}{2}\textnormal{-C}}

\allowdisplaybreaks

\title{A ZX-Calculus with Triangles for Toffoli-Hadamard, Clifford+T, and Beyond} 
\author{Renaud Vilmart
\institute{Universit\'e de Lorraine, CNRS, Inria, LORIA, F 54000 Nancy, France}
\email{renaud.vilmart@loria.fr}
}
\def\authorrunning{R. Vilmart}
\def\titlerunning{A ZX-Calculus for Toffoli-Hadamard}
\begin{document}

\maketitle
\begin{abstract}
We consider a ZX-calculus augmented with triangle nodes which is well-suited to reason on the so-called Toffoli-Hadamard fragment of quantum mechanics. We precisely show the form of the matrices it represents, and we provide an axiomatisation which makes the language complete for the Toffoli-Hadamard quantum mechanics. We extend the language with arbitrary angles and show that any true equation involving linear diagrams which constant angles are multiple of $\pi$ are derivable. We show that a single axiom is then necessary and sufficient to make the language equivalent to the ZX-calculus which is known to be complete for Clifford+T quantum mechanics. As a by-product, it leads to a new and simple complete axiomatisation for Clifford+T quantum mechanics. 

\end{abstract}

\section{Introduction}

The ZX-Calculus is a powerful graphical language for quantum computing \cite{picturing-qp}. It has been introduced in 2008 by Coecke and Duncan \cite{interacting}, and has several applications in quantum information processing (e.g. MBQC \cite{mbqc}, quantum codes \cite{chancellor2016coherent,de2017zx,duncan2014verifying,horsman2011quantum}).

The language manipulates diagrams, generated by roughly three kinds of vertices: \scalebox{0.7}{
\InputIfFileExists{gn-alpha-small.tikz}{}{\input{./figures/gn-alpha-small.tikz}}
}, \scalebox{0.7}{
\InputIfFileExists{rn-alpha-small.tikz}{}{\input{./figures/rn-alpha-small.tikz}}
} and \scalebox{0.7}{
\InputIfFileExists{Hadamard-small.tikz}{}{\input{./figures/Hadamard-small.tikz}}
}; and which represent quantum evolutions thanks to the standard interpretation. It comes with a set of allowed transformations that preserve the represented process. These rules can be implemented in the interactive proof assistant Quantomatic \cite{quanto,kissinger2015quantomatic}. The property that they preserve the semantics is called \emph{soundness}, and is easily verifiable. Its converse, \emph{completeness}, is obtained when any two diagrams that represent the same process can be transformed into one-another with only the rules of the ZX-Calculus.

The first completeness result was for a restriction of the language called Clifford \cite{pi-2-complete}. This result was adapted to a smaller restriction of the language, the so-called real stabilizer \cite{pivoting}. Problem was, these two restrictions are not universal: their diagrams cannot approximate all arbitrary evolution, and they are efficiently simulable on a classical computer. The ``easiest'' restriction of the language with approximate universality is the Clifford+T fragment. A first step was made when completeness was discovered for the 1-qubit operations of the restriction \cite{pi-4-single-qubit}. A complete axiomatisation for the many-qubits Clifford+T fragment was finally discovered in \cite{JPV}, and lead to a complete axiomatisation in the general case, first in \cite{NgWang,HNW} where two new generators were added, and then in \cite{JPV-universal} with the original syntax.

Thanks to its proximity to quantum circuitry, one of the potential uses of the ZX-Calculus is circuit simplification. The usual ZX-Calculus \cite{interacting} can be seen as an generalisation of quantum circuits built with the gate set (CNot, H, $R_Z(\alpha)$) where $\alpha$ may take any value in $\mathbb{R}$. Then, the real stabilizer ZX-Calculus corresponds to the set of gates when the angles are multiples of $\pi$, the Clifford restriction to when angles are multiples of $\frac{\pi}{2}$, and the Clifford+T restriction to when angles are multiples of $\frac{\pi}{4}$.

While the Clifford+T fragment (where T$=R_Z(\pi/4)$) of quantum mechanics is preeminent in quantum information processing, Toffoli-Hadamard quantum mechanics is also well known and widely used.  The Toffoli-Hadamard quantum mechanics is the real counter-part of the Clifford+T quantum mechanics: the matrices of the Toffoli+Hadamard fragment are exactly the real matrices of the Clifford+T fragment. Toffoli-Hadamard quantum mechanics is also known to be approximately universal for quantum computing \cite{toffoli-simple,toffoli}. The Toffoli gate is a 3-qubit unitary evolution which is nothing but a controlled CNot: it maps $\ket{x,y,z}$ to $\ket{x,y,x\wedge y\oplus z}$.  The Toffoli gate can be considered a generator of the considered language \cite{category-tof}, or can be decomposed into CNot, H, and T gates \cite{nielsen-chuang-2010,clifford+t} and hence can be represented in the \frag4 of the ZX-calculus. However, as pointed out in \cite{JPV,NgWang-clifford+t}, the Toffoli gate admits simpler representations using the triangle node:

%
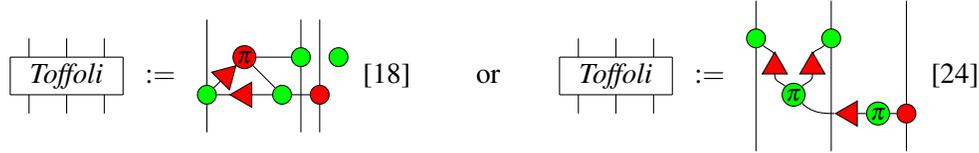
\begin{center}
\def\fig{toffoli-from-triangle}
$\input{./figures/\fig/\fig_00.tikz}~~:=~~\input{./figures/\fig/\fig_01.tikz}$~~\cite{JPV} $\qquad\text{or}\qquad\input{./figures/\fig/\fig_00.tikz}~~:=~~
\InputIfFileExists{toffoli-from-triangle-2.tikz}{}{\input{./figures/toffoli-from-triangle-2.tikz}}
$~~\cite{NgWang-clifford+t}
\end{center}
and the triangle itself can be recovered from the Toffoli gate:
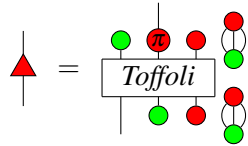
\def\fig{triangle-from-toffoli}
\[\input{./figures/\fig/\fig_00.tikz}\eq{}\input{./figures/\fig/\fig_01.tikz}\]

The triangle itself was introduced in \cite{JPV} as a notation for a diagram of the Clifford+T restriction, and was used as one of the additional generators in \cite{NgWang,NgWang-clifford+t,HNW}. Since the CNOT gate is depicted in ZX as the composition of two of its generators, we propose to add the triangle node as a generator for a version of the ZX-Calculus devoted to represent specifically Toffoli, without having to decompose it as a Clifford+T diagram.

We first present the diagrams of the ZX-Calculus augmented with the triangle node, and give an axiomatisation for the Toffoli-Hadamard quantum mechanics in Section \ref{sec:ZX-triangles}. We prove it is universal, sound and complete in Section \ref{sec:universal-sound-complete}. In Section \ref{sec:beyond-TH}, we try and adapt a theorem from \cite{JPV-universal} which gives a completeness result on a broader restriction of the language. We finally give in Section \ref{sec:clifford+t-triangles} a simple axiomatisation for the Clifford+T fragment of the ZX-Calculus with triangles (that corresponds to the (Toffoli, $H$, $R_Z(\pi/4)$) gate set in circuitry), and prove it is complete.

\section{\dzx-Calculus: A ZX-Calculus with Triangles}
\label{sec:ZX-triangles}

\subsection{Diagrams and Standard Interpretation}

A \dzx-diagram $D:k\to l$ with $k$ inputs and $l$ outputs is generated by:\\
\begin{center}
\bgroup
\def\arraystretch{2.5}
{\begin{tabular}{|@{~~}cc|@{~~}cc|@{~~~}cc@{~~}|}
\hline
$R_Z^{(n,m)}(\alpha):n\to m$ & 
\InputIfFileExists{gn-alpha-2.tikz}{}{\input{./figures/gn-alpha-2.tikz}}
 & 
$R_X^{(n,m)}(\alpha):n\to m$ & 
\InputIfFileExists{rn-alpha-2.tikz}{}{\input{./figures/rn-alpha-2.tikz}}
 & 
$H:1\to 1$ & 
\InputIfFileExists{Hadamard.tikz}{}{\input{./figures/Hadamard.tikz}}
\\\hline
$e:0\to 0$ & 
\InputIfFileExists{empty-diagram.tikz}{}{\input{./figures/empty-diagram.tikz}}
 &
$\mathbb{I}:1\to 1$ & 
\InputIfFileExists{single-line.tikz}{}{\input{./figures/single-line.tikz}}
 & 
$\sigma:2\to 2$ & 
\InputIfFileExists{crossing.tikz}{}{\input{./figures/crossing.tikz}}
\\\hline
$\epsilon:2\to 0$ & 
\InputIfFileExists{cup.tikz}{}{\input{./figures/cup.tikz}}
 & 
$\eta:0\to 2$ & 
\InputIfFileExists{caps.tikz}{}{\input{./figures/caps.tikz}}
 & 
$\mathrm{\Delta}:1\to 1$ & 
\InputIfFileExists{triangle.tikz}{}{\input{./figures/triangle.tikz}}
\\\hline
\end{tabular}}
\egroup\\
where $n,m\in \mathbb{N}$ and $\alpha \in \mathbb{R}$. The generator $e$ is the empty diagram.
\end{center}

\vspace{0.2cm}
and the two compositions:
\begin{itemize}
\item Spatial Composition: for any $D_1:a\to b$ and $D_2:c\to d$, $D_1\otimes D_2:a+c\to b+d$ consists in placing $D_1$ and $D_2$ side by side, $D_2$ on the right of $D_1$.
\item Sequential Composition: for any $D_1:a\to b$ and $D_2:b\to c$, $D_2\circ D_1:a\to c$ consists in placing $D_1$ on the top of $D_2$, connecting the outputs of $D_1$ to the inputs of $D_2$.
\end{itemize}

The standard interpretation of the \dzx-diagrams associates to any diagram $D:n\to m$ a linear map $\interp{D}:\mathbb{C}^{2^n}\to\mathbb{C}^{2^m}$ inductively defined as follows:\\
\needspace{2em}\noindent\titlerule{$\interp{.}$}
\[ \interp{D_1\otimes D_2}:=\interp{D_1}\otimes\interp{D_2} \qquad 
\interp{D_2\circ D_1}:=\interp{D_2}\circ\interp{D_1}\]
\[\interp{
\InputIfFileExists{empty-diagram.tikz}{}{\input{./figures/empty-diagram.tikz}}
~}:=\begin{pmatrix}
1
\end{pmatrix} \qquad
\interp{~
\InputIfFileExists{single-line.tikz}{}{\input{./figures/single-line.tikz}}
~~}:= \begin{pmatrix}
1 & 0 \\ 0 & 1\end{pmatrix}\qquad
\interp{~
\InputIfFileExists{Hadamard.tikz}{}{\input{./figures/Hadamard.tikz}}
~}:= \frac{1}{\sqrt{2}}\begin{pmatrix}1 & 1\\1 & -1\end{pmatrix}\qquad
\interp{
\InputIfFileExists{triangle.tikz}{}{\input{./figures/triangle.tikz}}
}=\begin{pmatrix}1&1\\0&1\end{pmatrix}\]
$$\interp{
\InputIfFileExists{crossing.tikz}{}{\input{./figures/crossing.tikz}}
}:= \begin{pmatrix}
1&0&0&0\\
0&0&1&0\\
0&1&0&0\\
0&0&0&1
\end{pmatrix} \qquad
\interp{\raisebox{-0.25em}{$
\InputIfFileExists{cup.tikz}{}{\input{./figures/cup.tikz}}
$}}:= \begin{pmatrix}
1&0&0&1
\end{pmatrix} \qquad
\interp{\raisebox{-0.35em}{$
\InputIfFileExists{caps.tikz}{}{\input{./figures/caps.tikz}}
$}}:= \begin{pmatrix}
1\\0\\0\\1
\end{pmatrix}$$
For any $\alpha\in\mathbb{R}$, $\interp{\begin{tikzpicture}
	\begin{pgfonlayer}{nodelayer}
		\node [style=gn] (0) at (0, -0) {$\alpha$};
	\end{pgfonlayer}
\end{tikzpicture}}:=\begin{pmatrix}1+e^{i\alpha}\end{pmatrix}$, and  for any $n,m\geq 0$ such that $n+m>0$:
$$
\interp{
\InputIfFileExists{gn-alpha.tikz}{}{\input{./figures/gn-alpha.tikz}}
}:=
\annoted{2^m}{2^n}{\begin{pmatrix}
  1 & 0 & \cdots & 0 & 0 \\
  0 & 0 & \cdots & 0 & 0 \\
  \vdots & \vdots & \ddots & \vdots & \vdots \\
  0 & 0 & \cdots & 0 & 0 \\
  0 & 0 & \cdots & 0 & e^{i\alpha}
 \end{pmatrix}}
\qquad\qquad
\interp{
\InputIfFileExists{rn-alpha.tikz}{}{\input{./figures/rn-alpha.tikz}}
}:=\interp{~
\InputIfFileExists{Hadamard.tikz}{}{\input{./figures/Hadamard.tikz}}
~}^{\otimes m}\circ \interp{
\InputIfFileExists{gn-alpha.tikz}{}{\input{./figures/gn-alpha.tikz}}
}\circ \interp{~
\InputIfFileExists{Hadamard.tikz}{}{\input{./figures/Hadamard.tikz}}
~}^{\otimes n}$$
\begin{minipage}{\columnwidth}
$\left(\text{where }M^{\otimes 0}=\begin{pmatrix}1\end{pmatrix}\text{ and }M^{\otimes k}=M\otimes M^{\otimes k-1}\text{ for any }k\in \mathbb{N}^*\right)$.\\
\rule{\columnwidth}{0.5pt}
\end{minipage}\\

To simplify, the red and green nodes will be represented empty when holding a 0 angle:
\[ 
\InputIfFileExists{gn-empty-is-gn-zero.tikz}{}{\input{./figures/gn-empty-is-gn-zero.tikz}}
 \qquad\text{and}\qquad 
\InputIfFileExists{rn-empty-is-rn-zero.tikz}{}{\input{./figures/rn-empty-is-rn-zero.tikz}}
 \]

\subsection{Calculus}

\dzx represents the general ZX-Calculus with triangle, i.e.~where the angles can take any value in $\mathbb{R}$. In the following, we will focus on some restrictions of the language. We call the \frag q of the \dzx, the restriction of the ZX-Calculus with triangles, where all the angles are multiples of $\frac{\pi}{q}$, and we denote it $\dzx_{\frac{\pi}{q}}$.

\begin{figure*}[!htb]
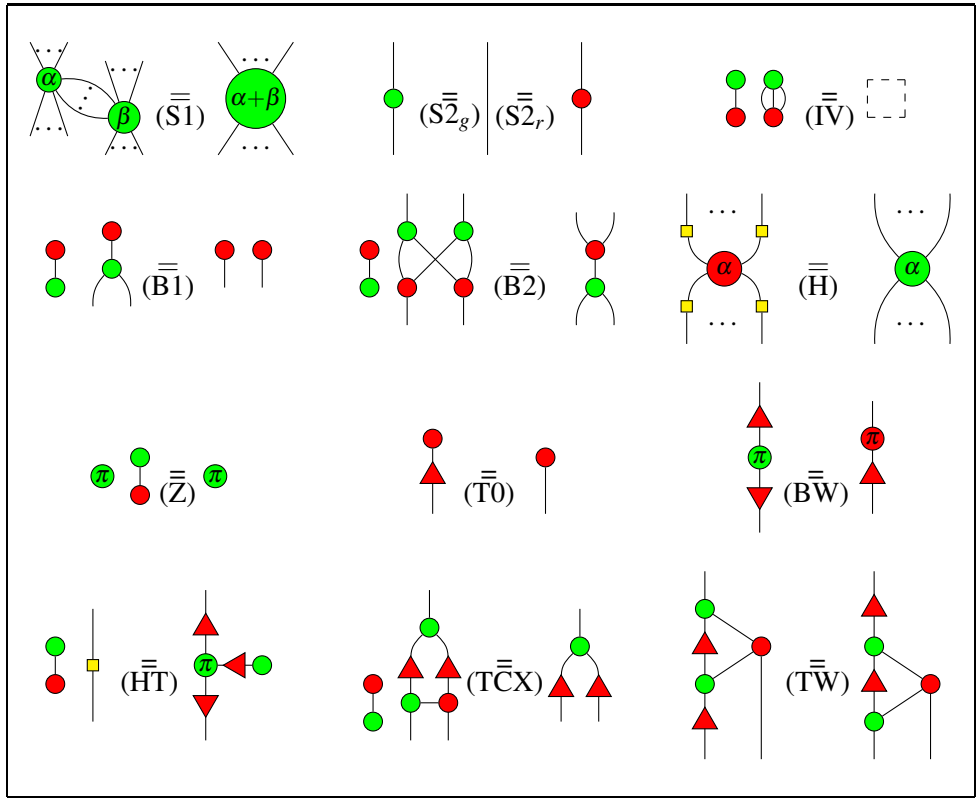

 \centering
 \hypertarget{r:rules}{}
 \begin{tabular}{|c@{$\qquad$}c@{$\qquad$}c|}
   \hline
   && \\
   
\InputIfFileExists{spider-1.tikz}{}{\input{./figures/spider-1.tikz}}
&
\InputIfFileExists{s2-green-red.tikz}{}{\input{./figures/s2-green-red.tikz}}
&
\InputIfFileExists{inverse.tikz}{}{\input{./figures/inverse.tikz}}
\\
   && \\
   
\InputIfFileExists{b1s.tikz}{}{\input{./figures/b1s.tikz}}
& 
\InputIfFileExists{b2s.tikz}{}{\input{./figures/b2s.tikz}}
&
\InputIfFileExists{h2.tikz}{}{\input{./figures/h2.tikz}}
\\
   && \\
   
\InputIfFileExists{zero-rule.tikz}{}{\input{./figures/zero-rule.tikz}}
&
\InputIfFileExists{ket-0-on-triangle.tikz}{}{\input{./figures/ket-0-on-triangle.tikz}}
&
\InputIfFileExists{BW.tikz}{}{\input{./figures/BW.tikz}}
\\
   && \\
   
\InputIfFileExists{triangle-hadamard.tikz}{}{\input{./figures/triangle-hadamard.tikz}}
&
\InputIfFileExists{cnot-on-triangle-fork.tikz}{}{\input{./figures/cnot-on-triangle-fork.tikz}}
&
\InputIfFileExists{triangle-through-W.tikz}{}{\input{./figures/triangle-through-W.tikz}}
\\
   && \\
   \hline
  \end{tabular}
 \caption[]{Set of rules $\dzx$ for the ZX-Calculus with triangles. The right-hand side of (IV) is an empty diagram. (...) denote zero or more wires, while (\protect\rotatebox{45}{\raisebox{-0.4em}{$\cdots$}}) denote one or more wires. $\alpha,\beta\in\mathbb{R}$. $\dzx_{\pi}$ is obtained when restricting the angles $\alpha,\beta\in\{0,\pi\}$.
 }
 \label{fig:ZX-rules}
\end{figure*}

Two different diagrams may represent the same quantum evolution i.e.~there exist two different diagrams $D_1$ and $D_2$ such that $\interp{D_1}=\interp{D_2}$. This issue is addressed by giving a set of axioms: a set of permitted local diagram transformations that preserve the semantics. We give in Figure \ref{fig:ZX-rules} a set of axioms for the general ZX-Calculus with triangles, and we define the $\dzx_{\pi}$ as the same set of axioms where the angles are restricted to $\{0,\pi\}$.

\begin{remark}
In the following, we will freely use the notation $\dzx_{\frac{\pi}{q}}$ to denote either the set of diagrams in the \frag q or the set of rules given for these specific diagrams.
\end{remark}

This set of axioms consists of the rules for the real stabiliser ZX-Calculus given in \cite{pivoting} -- except the so-called H-Loop: \scalebox{0.8}{\def\fig{h-loop-from-v2}$\input{./figures/\fig/\fig_00.tikz}\eq{}\input{./figures/\fig/\fig_07.tikz}$} --, augmented with 5 rules that include the node $\mathrm{\Delta}$. However, this missing axiom can be retrieved (Lemma \ref{lem:h-loop}), so all the equations of the $\pi$-fragment that do not involve $\mathrm{\Delta}$-nodes are derivable.

Additionally to these rules, the paradigm ``Only Topology Matters'' ensures we can be as lax as we want when manipulating diagrams, in the sense that what only matters is if two nodes are connected or not (and how many times). A distinction has to be made for $\mathrm{\Delta}$ where the orientation of the node matters. For instance:
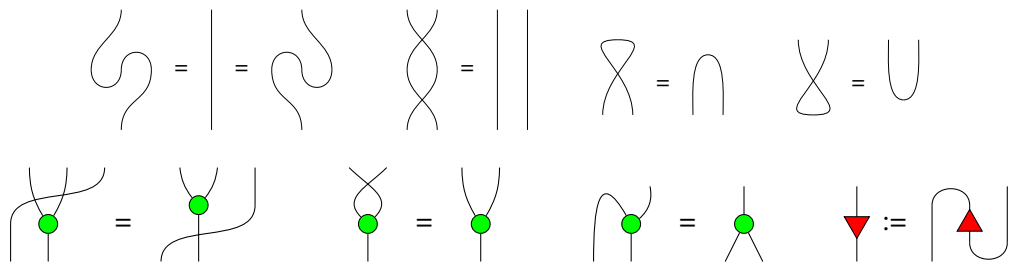
\[\scalebox{0.8}{
\InputIfFileExists{bent-wire.tikz}{}{\input{./figures/bent-wire.tikz}}
}\]
\[{
\InputIfFileExists{bent-wire-2.tikz}{}{\input{./figures/bent-wire-2.tikz}}
}\]
If we can transform a diagram $D_1$ into a diagram $D_2$ using the rules of \dzx or the previous paradigm, we write $\dzx\vdash D_1=D_2$. We said earlier that ``the rules can be locally applied''. This means that for any three diagrams, $D_1, D_2$, and $D$, if $\dzx\vdash D_1=D_2$, then:\\
\hspace*{2em}\begin{tabular}{@{$\bullet\quad$}l@{$\qquad\qquad\bullet\quad$}l}
$\dzx\vdash D_1\circ D = D_2\circ D$&
$\dzx\vdash D\circ D_1 = D\circ D_2$\\
$\dzx\vdash D_1\otimes D = D_2\otimes D$&
$\dzx\vdash D\otimes D_1 = D\otimes D_2$
\end{tabular}

An important property of the usual ZX-Calculus is that colour-swapping the diagrams (transforming green nodes in red nodes and vice-versa) preserves the equality. We made the triangle red, and we can define a green triangle as $
\InputIfFileExists{green-triangle-definition.tikz}{}{\input{./figures/green-triangle-definition.tikz}}
$ so that this property is preserved.

\section{Universality, Soundness, Completeness}
\label{sec:universal-sound-complete}
The Toffoli+Hadamard gate set in circuitry is (approximately) universal \cite{toffoli-simple,toffoli}: any quantum evolution can be approximated with arbitrary precision by a circuit with this set of gates. It is expected that the result extends to the adequate fragment of the \dzx-Calculus. It does:

\begin{theorem}
\label{thm:universal}
For any matrix $M\in\frac{1}{\sqrt{2}^{\mathbb{N}}}\mathcal{M}_{2^m,2^n}(\mathbb{Z})$, there exists a $\dzx_{\pi}$-diagram $D:n\to m$ such that $M=\interp{D}$.
\end{theorem}
In other words, the fragment represents exactly the integer matrices of dimensions powers of two and multiplied by an arbitrary power of $\frac{1}{\sqrt{2}}$. One inclusion is easy to notice: the diagrams represent matrices in $\frac{1}{\sqrt{2}^{\mathbb{N}}}\mathcal{M}_{2^m,2^n}(\mathbb{Z})$. Indeed, the standard interpretation of all the generators is an integer matrix, multiplied by $\frac{1}{\sqrt{2}}$ in the case of the Hadamard gate, and the two compositions preserve this structure. The second inclusion, the fact that any matrix can be represented as a $\dzx_{\pi}$-diagram is less trivial, and will be shown in the following.

The language is not only universal, it is \emph{sound}: the transformations do preserve the represented quantum evolution i.e.~$\dzx_{\pi}\vdash D_1=D_2 \implies \interp{D_1}=\interp{D_2}$. This is a routine check. The converse, however is harder to prove. \emph{Completeness} is obtained when two diagrams can be transformed into one another whenever they represent the same evolution. This is our main theorem:

\begin{theorem}
\label{thm:complete}
The set of rules in Figure \ref{fig:ZX-rules} makes the $\pi$-fragment of the \dzx-Calculus complete. For any diagrams $D_1$ and $D_2$ of the fragment,
\[\interp{D_1}=\interp{D_2} \iff \dzx_{\pi}\vdash D_1=D_2\]
\end{theorem}

The proof uses the now usual method of a back and forth interpretation from the ZX-Calculus to the ZW-Calculus, another graphical language suited for quantum processes.

\subsection{ZW-Calculus}

The ZW-Calculus, introduced in 2010 as the GHZ/W-calculus by Coecke and Kissinger \cite{ghz-w}, was completed by Hadzihasanovic in 2015 \cite{zw}. This first complete language is only universal for matrices over $\mathbb{Z}$.
Even though it has been extended to represent any complex matrix by the last author \cite{Amar}, the first version, closer to the expressive power of the $\pi$-fragment of the \dzx-Calculus, is the one we will use in the following of the paper.

The ZW-diagrams are generated by:
\[T_e=\left\lbrace 

\InputIfFileExists{Z-1-1.tikz}{}{\input{./figures/Z-1-1.tikz}}
~,~
\InputIfFileExists{Z-2-1.tikz}{}{\input{./figures/Z-2-1.tikz}}
~,~
\InputIfFileExists{W-1-1.tikz}{}{\input{./figures/W-1-1.tikz}}
~,~
\InputIfFileExists{W-1-2.tikz}{}{\input{./figures/W-1-2.tikz}}
~,\scalebox{1}{
\InputIfFileExists{single-line.tikz}{}{\input{./figures/single-line.tikz}}
}~,\raisebox{-0.3em}{
\InputIfFileExists{cup.tikz}{}{\input{./figures/cup.tikz}}
}~,\raisebox{-0.4em}{
\InputIfFileExists{caps.tikz}{}{\input{./figures/caps.tikz}}
}~,~
\InputIfFileExists{crossing.tikz}{}{\input{./figures/crossing.tikz}}
~,~
\InputIfFileExists{zw-cross.tikz}{}{\input{./figures/zw-cross.tikz}}
~,~
\InputIfFileExists{empty-diagram.tikz}{}{\input{./figures/empty-diagram.tikz}}
~
\right\rbrace\]
The generators are then composed using the two same -- sequential and spatial -- compositions.

Again, the diagrams represent quantum evolutions, so the language comes with a standard interpretation, inductively defined as:\\
\needspace{2em}\noindent\titlerule{$\interp{.}$}
$$ \interp{D_1\otimes D_2}:=\interp{D_1}\otimes\interp{D_2} \qquad 
\interp{D_2\circ D_1}:=\interp{D_2}\circ\interp{D_1}\qquad
\interp{
\InputIfFileExists{empty-diagram.tikz}{}{\input{./figures/empty-diagram.tikz}}
~}:=\begin{pmatrix}1\end{pmatrix} \quad
\interp{~
\InputIfFileExists{single-line.tikz}{}{\input{./figures/single-line.tikz}}
~~}:= \begin{pmatrix}
1 & 0 \\ 0 & 1\end{pmatrix}$$
$$ 
\interp{
\InputIfFileExists{crossing.tikz}{}{\input{./figures/crossing.tikz}}
}:= \begin{pmatrix}
1&0&0&0\\
0&0&1&0\\
0&1&0&0\\
0&0&0&1
\end{pmatrix} \qquad
\interp{
\InputIfFileExists{zw-cross.tikz}{}{\input{./figures/zw-cross.tikz}}
}:= \begin{pmatrix}
1&0&0&0\\
0&0&1&0\\
0&1&0&0\\
0&0&0&-1
\end{pmatrix}\qquad
\interp{\raisebox{-0.4em}{$
\InputIfFileExists{caps.tikz}{}{\input{./figures/caps.tikz}}
$}}:= \begin{pmatrix}
1\\0\\0\\1
\end{pmatrix}\qquad
\interp{\raisebox{-0.3em}{$
\InputIfFileExists{cup.tikz}{}{\input{./figures/cup.tikz}}
$}}:= \begin{pmatrix}
1\\0\\0\\1
\end{pmatrix}^t$$
$$
\interp{
\InputIfFileExists{W-1-1.tikz}{}{\input{./figures/W-1-1.tikz}}
}:= \begin{pmatrix}
0&1\\1&0
\end{pmatrix} \qquad 
\interp{
\InputIfFileExists{W-1-2.tikz}{}{\input{./figures/W-1-2.tikz}}
}:= \begin{pmatrix}
0&1\\1&0\\1&0\\0&0
\end{pmatrix}\qquad
\interp{
\InputIfFileExists{Z-1-1.tikz}{}{\input{./figures/Z-1-1.tikz}}
}:= \begin{pmatrix}
1&0\\0&-1
\end{pmatrix} \qquad 
\interp{
\InputIfFileExists{Z-2-1.tikz}{}{\input{./figures/Z-2-1.tikz}}
}:= \begin{pmatrix}
1&0&0&0\\0&0&0&-1
\end{pmatrix}
$$
\rule{\columnwidth}{0.5pt}

\begin{figure*}[!bt]
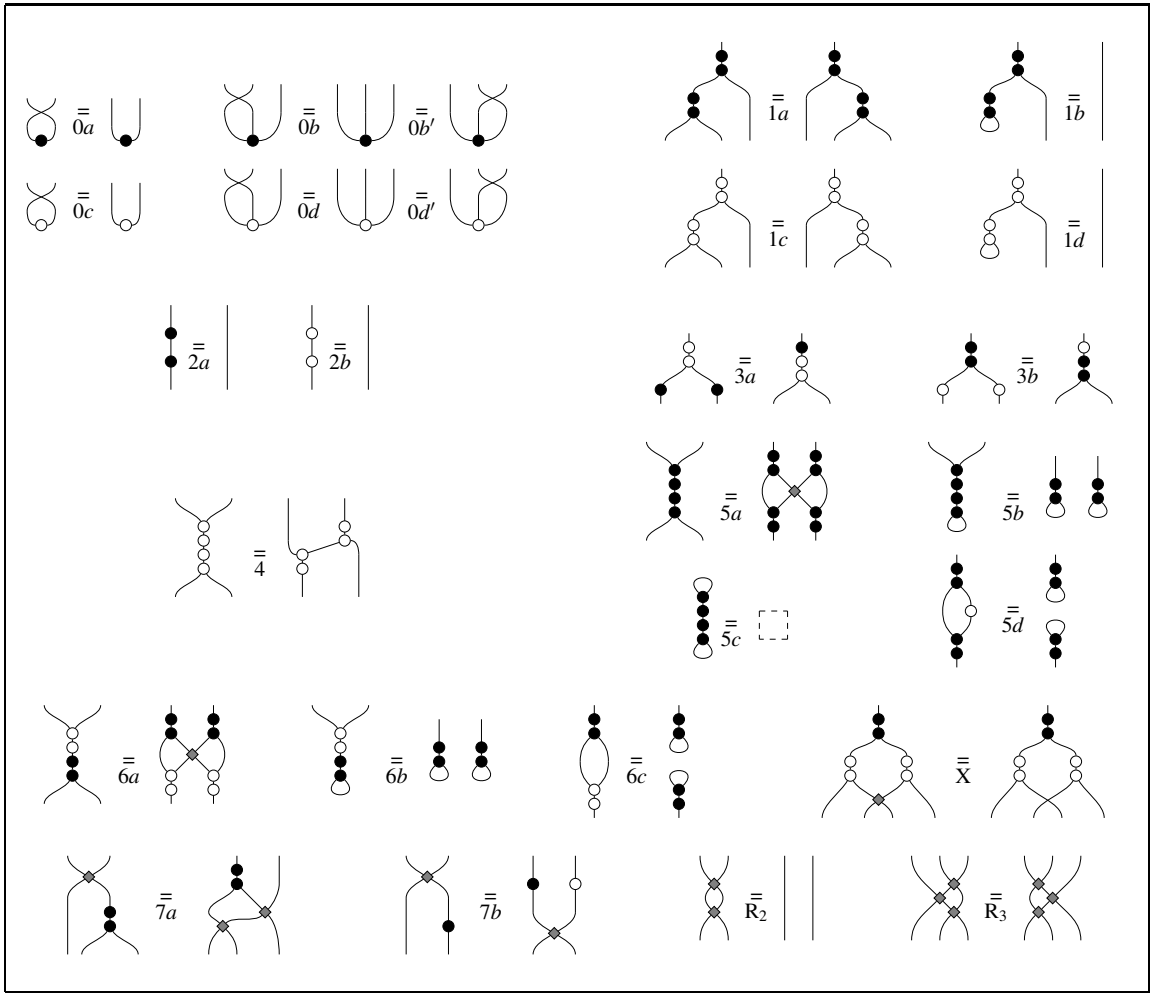

\def\scale{0.75}
\centering
\begin{tabular}{|ccc|}
\hline
&&\\
\scalebox{\scale}{
\InputIfFileExists{ZW-rule-0-no-braid.tikz}{}{\input{./figures/ZW-rule-0-no-braid.tikz}}
} & $\qquad$ & \scalebox{\scale}{
\InputIfFileExists{ZW-rule-1.tikz}{}{\input{./figures/ZW-rule-1.tikz}}
} \\
&&\\
\scalebox{\scale}{
\InputIfFileExists{ZW-rule-2.tikz}{}{\input{./figures/ZW-rule-2.tikz}}
} && \scalebox{\scale}{
\InputIfFileExists{ZW-rule-3.tikz}{}{\input{./figures/ZW-rule-3.tikz}}
} \\
&&\\
 \scalebox{\scale}{
\InputIfFileExists{ZW-rule-4.tikz}{}{\input{./figures/ZW-rule-4.tikz}}
} && \scalebox{\scale}{
\InputIfFileExists{ZW-rule-5-no-braid.tikz}{}{\input{./figures/ZW-rule-5-no-braid.tikz}}
} \\
 &&\\
 \multicolumn{3}{|c|}{\scalebox{\scale}{
\InputIfFileExists{ZW-rule-6-no-braid.tikz}{}{\input{./figures/ZW-rule-6-no-braid.tikz}}
} $\qquad\quad$ \scalebox{\scale}{
\InputIfFileExists{ZW-rule-X-no-braid.tikz}{}{\input{./figures/ZW-rule-X-no-braid.tikz}}
}}\\
 &&\\
 \multicolumn{3}{|c|}{
 \scalebox{\scale}{
\InputIfFileExists{ZW-rule-7-no-braid.tikz}{}{\input{./figures/ZW-rule-7-no-braid.tikz}}
} $\qquad\quad$ \scalebox{\scale}{
\InputIfFileExists{reidmeister.tikz}{}{\input{./figures/reidmeister.tikz}}
}} \\
 &&\\
\hline
\end{tabular}
\caption{Set of rules for the ZW-Calculus.}
 \label{fig:ZW-rules}
\end{figure*}

The language also comes with its own set of axioms, given in Figure \ref{fig:ZW-rules}.
Here again, the paradigm ``Only Topology Matters'' is respected (except for 
\InputIfFileExists{zw-cross.tikz}{}{\input{./figures/zw-cross.tikz}}
 where the ordering of inputs and outputs is not to be messed with). Using this, we define 1-legged black and white dots as 3-legged dots with a loop: 
\InputIfFileExists{black-dot-0-1.tikz}{}{\input{./figures/black-dot-0-1.tikz}}
 and 
\InputIfFileExists{white-dot-0-1.tikz}{}{\input{./figures/white-dot-0-1.tikz}}
, and we can build any 2-legged or 3-legged white and black dots. For instance: 
\InputIfFileExists{zw-bent-wire-example.tikz}{}{\input{./figures/zw-bent-wire-example.tikz}}


The language is of course sound, and as previously said, complete.

\subsection{\zwh-Calculus}

The $\pi$-fragment of the \dzx-Calculus represents elements of $\frac{1}{\sqrt{2}^\mathbb{N}}\mathcal{M}_{2^m,2^n}\left(\mathbb{Z}\right)$, while the ZW-Cal\-culus represents elements in $\mathcal{M}_{2^m,2^n}\left(\mathbb{Z}\right)$. If the two languages had the same expressive power, it would greatly simplify the interpretation from one another, by avoiding complicated encodings such as in \cite{JPV}.

What is missing from the ZW-Calculus is merely a global scalar, which can be any power of $\frac{1}{\sqrt{2}}$. Intuitively, adding a generator, which only represents the scalar $\frac{1}{\sqrt{2}}$ should be enough, for tensoring it to a diagram $D$ should yield $\frac{1}{\sqrt{2}}\interp{D}$, and the different powers should be obtained by putting side by side enough occurrences of the new scalar. This should deal with the expressive power of the language. It remains then to bind the new generator to the others in the axiomatisation.

\begin{definition}
We define the \zwh-Calculus as the extension of the ZW-Calculus such as:
\[\left\lbrace
\begin{array}{l}
 T_{\!\!\frac 1{\sqrt{2}}}=T_e\cup\{\half\}\\
\zwh=\zw\cup\left\lbrace
\InputIfFileExists{additional-ZW-rule.tikz}{}{\input{./figures/additional-ZW-rule.tikz}}
~,~~~
\InputIfFileExists{additional-ZW-rule-zero.tikz}{}{\input{./figures/additional-ZW-rule-zero.tikz}}
\right\rbrace
\end{array}
\right.\]
The standard interpretation of a diagram $D:n\to m$ is now a matrix
$\interp{D}\in\frac{1}{\sqrt{2}^\mathbb{N}}\mathcal{M}_{2^m,2^n}\left(\mathbb{Z}\right)$ that is an integer matrix times a power of $\frac{1}{\sqrt{2}}$, and is given by
the standard interpretation of the ZW-Calculus extended with $\interp{\half}:=\frac{1}{\sqrt{2}}$.
\end{definition}

By building on top of the already complete ZW-Calculus, some results are preserved or easily extendable. For instance, knowing that the ZW-Calculus is sound, showing that the rules $iv$ and $z$ are sound is enough to conclude that the \zwh-Calculus is sound. For the completeness:

\begin{proposition}
  \label{prop:zwcomplete}
  The \zwh is sound and complete:
  For two diagrams $D_1, D_2$ of the \zwh-calculus,
  $$\interp{D_1} = \interp{D_2}\iff\zwh \vdash D_1 = D_2$$
\end{proposition}

We now have a complete version of the ZW-Calculus, with the same -- yet to be shown -- expressive power as the $\pi$-fragment of the \dzx-Calculus.

\subsection{From \zwh to \dzx, and Expressive Power of the Latter}

We define here an interpretation $[.]_X$ that transforms any diagram of the \zwh-Calculus into a \dzx-diagram:\\
\needspace{6em}\noindent\titlerule{$[.]_X$}
\begin{multicols}{3}
\[ 
\InputIfFileExists{empty-diagram.tikz}{}{\input{./figures/empty-diagram.tikz}}
 \quad\mapsto\quad 
\InputIfFileExists{empty-diagram.tikz}{}{\input{./figures/empty-diagram.tikz}}
\]\vfill
\[ 
\InputIfFileExists{single-line.tikz}{}{\input{./figures/single-line.tikz}}
 \quad\mapsto\quad 
\InputIfFileExists{single-line.tikz}{}{\input{./figures/single-line.tikz}}
\]\vfill
\[ 
\InputIfFileExists{caps.tikz}{}{\input{./figures/caps.tikz}}
 \quad \raisebox{0.3em}{$\mapsto$} \quad 
\InputIfFileExists{caps.tikz}{}{\input{./figures/caps.tikz}}
\]\vfill
\[ 
\InputIfFileExists{cup.tikz}{}{\input{./figures/cup.tikz}}
 \quad\raisebox{0.3em}{$\mapsto$}\quad 
\InputIfFileExists{cup.tikz}{}{\input{./figures/cup.tikz}}
\]\vfill
\[ 
\InputIfFileExists{crossing.tikz}{}{\input{./figures/crossing.tikz}}
 \quad\mapsto\quad 
\InputIfFileExists{crossing.tikz}{}{\input{./figures/crossing.tikz}}
\]\vfill
\[
\InputIfFileExists{ZW-to-ZX-cross.tikz}{}{\input{./figures/ZW-to-ZX-cross.tikz}}
\]\vfill
\[ \half \quad\mapsto\quad 
\InputIfFileExists{1_sqrt-2-ZX.tikz}{}{\input{./figures/1_sqrt-2-ZX.tikz}}
 \]\vfill
\[
\InputIfFileExists{ZW-to-ZX-white-dot-1-1.tikz}{}{\input{./figures/ZW-to-ZX-white-dot-1-1.tikz}}
\]\vfill
\[
\InputIfFileExists{ZW-to-ZX-white-dot-2-1.tikz}{}{\input{./figures/ZW-to-ZX-white-dot-2-1.tikz}}
\]\vfill
\[
\InputIfFileExists{ZW-to-ZX-dot-1-1.tikz}{}{\input{./figures/ZW-to-ZX-dot-1-1.tikz}}
\]
\[
\InputIfFileExists{ZW-to-ZX-dot-1-2-simplified.tikz}{}{\input{./figures/ZW-to-ZX-dot-1-2-simplified.tikz}}
\]
\end{multicols}

\noindent\begin{minipage}{\columnwidth}
\[D_1\circ D_2\mapsto [D_1]_X\circ [D_2]_X\qquad D_1\otimes D_2\mapsto [D_1]_X\otimes [D_2]_X\]
\rule{\columnwidth}{0.5pt}
\end{minipage}\\~\\
This interpretation preserves the semantics of the languages:
\begin{proposition}
\label{prop:WX-interp-semantics}
Let $D$ be a \zwh-diagram. Then $\interp{[D]_X}=\interp{D}$.
\end{proposition}
This is a routine check. The existence of this interpretation combined with the previous proposition is enough to prove Theorem \ref{thm:universal}:
\begin{proof}[Theorem \ref{thm:universal}]
Let $M\in\frac{1}{\sqrt{2}^\mathbb{N}}\mathcal{M}_{2^m,2^n}\left(\mathbb{Z}\right)$. There exist $n\in\mathbb{N}$ and $M'\in\mathcal{M}_{2^m,2^n}\left(\mathbb{Z}\right)$ such that $M=\frac{1}{\sqrt{2}^n}M'$. $M'$ is an integer matrix, so there exists $D_W'$ a ZW-diagram such that $\interp{D_W'}=M'$. We can then build the \zwh-diagram $D_W:=D_W'\otimes (\half)^{\otimes n}$ and notice that $\interp{D_W}=M$. Finally, we build the $\dzx_{\pi}$-diagram $D_X:=[D_W]_X$ and $\interp{D_X}=M$ by Proposition \ref{prop:WX-interp-semantics}.
\end{proof}

Another very important result is that the \dzx-Calculus proves the interpretation of all the rules of the \zwh-Calculus. More specifically:
\begin{proposition}
\label{prop:ZX-proves-ZW}
For any \zwh-diagrams $D_1$ and $D_2$:
\[\zwh\vdash D_1=D_2 \implies \dzx\vdash [D_1]_X=[D_2]_X\]
\end{proposition}
The proof is in appendix.

\subsection{From $\dzx_{\pi}$ to \zwh, and Completeness}

We now define an interpretation from the $\pi$-fragment of the \dzx-Calculus to the \zwh-Calculus.

\noindent\begin{minipage}{\columnwidth}
\titlerule{$[.]_W$}\\
\begin{multicols}{3}
\[
\InputIfFileExists{empty-diagram.tikz}{}{\input{./figures/empty-diagram.tikz}}
 \quad\mapsto\quad 
\InputIfFileExists{empty-diagram.tikz}{}{\input{./figures/empty-diagram.tikz}}
\]\vfill
\[
\InputIfFileExists{single-line.tikz}{}{\input{./figures/single-line.tikz}}
 \quad\mapsto\quad 
\InputIfFileExists{single-line.tikz}{}{\input{./figures/single-line.tikz}}
\]\vfill
\[
\InputIfFileExists{crossing.tikz}{}{\input{./figures/crossing.tikz}}
 \quad\mapsto\quad 
\InputIfFileExists{crossing.tikz}{}{\input{./figures/crossing.tikz}}
\]
\end{multicols}
\end{minipage}
\begin{multicols}{3}
\[
\InputIfFileExists{caps.tikz}{}{\input{./figures/caps.tikz}}
 \quad\raisebox{0.3em}{$\mapsto$}\quad 
\InputIfFileExists{caps.tikz}{}{\input{./figures/caps.tikz}}
\]\vfill
\[
\InputIfFileExists{cup.tikz}{}{\input{./figures/cup.tikz}}
 \quad\raisebox{0.3em}{$\mapsto$}\quad 
\InputIfFileExists{cup.tikz}{}{\input{./figures/cup.tikz}}
\]\vfill
\[
\InputIfFileExists{hadamard-ZX-to-ZW.tikz}{}{\input{./figures/hadamard-ZX-to-ZW.tikz}}
\]\vfill
\[
\InputIfFileExists{triangle-to-ZW.tikz}{}{\input{./figures/triangle-to-ZW.tikz}}
\]\vfill
\[
\InputIfFileExists{gn-to-ZW.tikz}{}{\input{./figures/gn-to-ZW.tikz}}
\]
\end{multicols}
$$\forall \alpha\in\{0,\pi\},\quad
\InputIfFileExists{rn-alpha.tikz}{}{\input{./figures/rn-alpha.tikz}}
\mapsto\left[~
\InputIfFileExists{Hadamard.tikz}{}{\input{./figures/Hadamard.tikz}}
~\right]_W^{\otimes m}\circ \left[
\InputIfFileExists{gn-alpha.tikz}{}{\input{./figures/gn-alpha.tikz}}
\right]_W\circ \left[~
\InputIfFileExists{Hadamard.tikz}{}{\input{./figures/Hadamard.tikz}}
~\right]_W^{\otimes n}
$$
\noindent\begin{minipage}{\columnwidth}
$$D_1\circ D_2\mapsto [D_1]_W\circ[D_2]_W\qquad\quad D_1\otimes D_2\mapsto [D_1]_W\otimes[D_2]_W$$
\rule{\columnwidth}{0.5pt}
\end{minipage}\\
Here, 
\InputIfFileExists{white-dot-1-1-power-p.tikz}{}{\input{./figures/white-dot-1-1-power-p.tikz}}
 either represents the identity if $p=0$, or the $1\to1$ white dot if $p=1$. The complexity of the interpretation of the green dot (and subsequently of the red dot), is merely due to the fact that we chose to present the \emph{expanded} version of the ZW-Calculus \cite{zw}, where spiders are ``exploded'' as compositions of 2- and 3-legged dots.

Again, the interpretation preserves the semantics:
\begin{proposition}
\label{prop:XW-interp-semantics}
Let $D$ be a $\dzx_{\pi}$-diagram. Then $\interp{[D]_W}=\interp{D}$.
\end{proposition}
And again, this is a routine check. Now, when composing the two interpretation, we can easily retrieve the initial diagram:
\begin{proposition}
\label{prop:double-interp}
For any diagram $D$ of the $\pi$-fragment of the \dzx-Calculus, $\dzx_{\pi}\vdash [[D]_W]_X = D$.
\end{proposition}

We can now prove the completeness theorem:
\begin{proof}[Theorem \ref{thm:complete}]
Let $D_1$ and $D_2$ be two diagrams of the $\dzx_{\pi}$ such that $\interp{D_1}=\interp{D_2}$. By Proposition \ref{prop:XW-interp-semantics}, $\interp{[D_1]_W}=\interp{[D_2]_W}$. By completeness of the \zwh (Proposition \ref{prop:zwcomplete}), $\zwh\vdash [D_1]_W=[D_2]_W$. By Proposition \ref{prop:ZX-proves-ZW}, $\dzx_{\pi}\vdash [[D_1]_W]_X=[[D_2]_W]_X$. Finally, by Proposition \ref{prop:double-interp}, $\dzx_{\pi}\vdash D_1=D_2$, hence proving the completeness of the $\dzx_{\pi}$-fragment.
\end{proof}

\section{Beyond Toffoli+Hadamard}
\label{sec:beyond-TH}

The set of axioms in Figure \ref{fig:ZX-rules} features two rules \so and \h, with parameters. In $\dzx_{\pi}$, they are  limited to $\{0,\pi\}$, but not in $\dzx$. We want to know what we can prove without the limitation to $\alpha,\beta\in\{0,\pi\}$ in the set of rules. To do so, we are going to adapt a result that was developed in \cite{JPV-universal}, and see what is missing to prove equalities that are valid for \emph{linear diagrams} with constants in $\{0,\pi\}$. The following of the section will give a sketch of the proof but will not dive into its details. For more, see \cite{JPV-universal}.

We call a \emph{linear diagram}, a diagram where some angles are treated as variables, and where the parameters of the green and red dots are affine combinations of the variables, where coefficients are in $\mathbb{Z}$:
\begin{definition}
A \dzx-diagram is linear in $\alpha_1, \ldots, \alpha_k$ with constants in $C\subseteq \mathbb R$, if it is generated by $R_Z^{(n,m)}(E)$, $R_X^{(n,m)}(E)$, $H$, $\mathrm{\Delta}$, $e$, $\mathbb I$, $\sigma$, $\epsilon$, $\eta$, and the spatial and sequential compositions, where $n,m\in \mathbb  N$, and $E$ is of the form $\sum_{i} n_i \alpha_i+c$, with $n_i\in \mathbb Z$ and $c\in C$. 
\end{definition}
For instance:
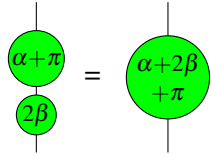
\[
\InputIfFileExists{linear-diagram-example.tikz}{}{\input{./figures/linear-diagram-example.tikz}}
\]
is a well-formed equation on linear diagrams, with constants in $\{0,\pi\}$. It is even sound, for it is sound for any values of $\alpha$ and $\beta$.

The idea of the proof is to separate the different occurrences of the variables from the rest of the diagram using the rule \so, change their colour if they are in red nodes using \h, and change the sign in front of the variables if needs be using the equation:
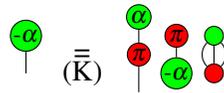
\[
\InputIfFileExists{K-bis.tikz}{}{\input{./figures/K-bis.tikz}}
\]
Then, the scalars with -$\alpha$ can be remove in favour of scalars with $\alpha$ on the other side of the equation, for, thanks to \iv and Lemma \ref{lem:multiplying-global-phases}: $\dzx\vdash \scalebox{0.8}{
\InputIfFileExists{scalar-e-pow-minus-i-alpha.tikz}{}{\input{./figures/scalar-e-pow-minus-i-alpha.tikz}}
}D_1=D_2 \equi{}\zx\vdash D_1=\scalebox{0.8}{
\InputIfFileExists{scalar-e-pow-i-alpha.tikz}{}{\input{./figures/scalar-e-pow-i-alpha.tikz}}
}D_2$.

First considering the case of a unique variable, it results that for any pair of diagrams $D_1(\alpha)$ and $D_2(\alpha)$ with variable $\alpha$, there exists a pair of variable-free diagrams $D_1'$ and $D_2'$ such that:
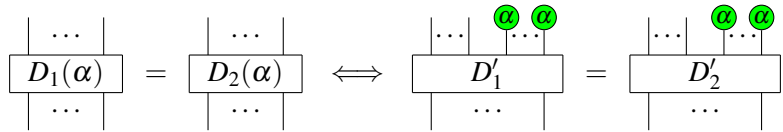
\def\fig{sketch-proof-provability-theorem}
\[\input{./figures/\fig/\fig_00.tikz}\eq{}\input{./figures/\fig/\fig_01.tikz}~~\iff~~\input{./figures/\fig/\fig_02.tikz}\eq{}\input{./figures/\fig/\fig_03.tikz}\]

We then have to find a diagram that is precisely a projector onto the span of $\left(
\InputIfFileExists{gn-alpha-0-1.tikz}{}{\input{./figures/gn-alpha-0-1.tikz}}
\right)^{\otimes r}$. We give the family $(P_r)$ of diagrams, inductively defined as:
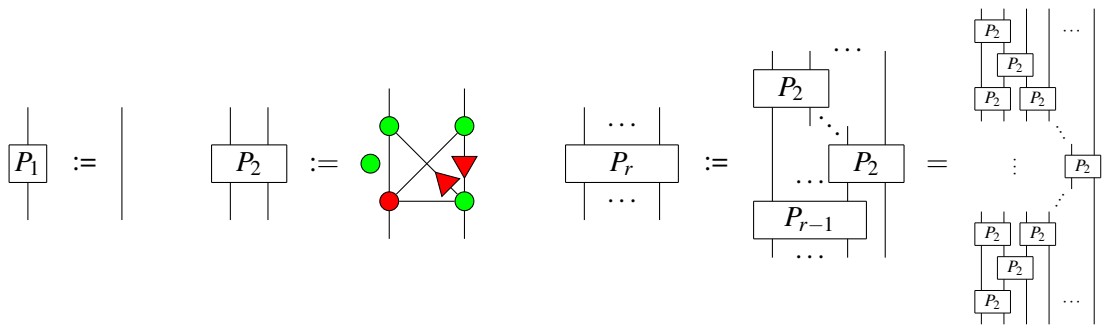
\[

\InputIfFileExists{P1.tikz}{}{\input{./figures/P1.tikz}}
\qquad\quad

\InputIfFileExists{2-in-2-out-box-M_2.tikz}{}{\input{./figures/2-in-2-out-box-M_2.tikz}}
~~:=~~ 
\InputIfFileExists{matrix-M2.tikz}{}{\input{./figures/matrix-M2.tikz}}
\qquad\quad

\InputIfFileExists{matrix-M-n-def.tikz}{}{\input{./figures/matrix-M-n-def.tikz}}
~~=~~\scalebox{0.6}{
\InputIfFileExists{matrix-M-n-form.tikz}{}{\input{./figures/matrix-M-n-form.tikz}}
}
\]
One can check that
$\interp{P_2} = \begin{pmatrix}
  1 & 0 & 0 & 0 \\
  0 & 0 & 1 & 0 \\
  0 & 0 & 1 & 0 \\
  0 & 0 & 0 & 1 \\
  \end{pmatrix}$ 
from which, thanks to \cite{JPV-universal}, we can immediately deduce that $P_r$ is a projector onto $\operatorname{span}\left\lbrace\interp{\left(
\InputIfFileExists{gn-alpha-0-1.tikz}{}{\input{./figures/gn-alpha-0-1.tikz}}
\right)^{\otimes r}}/\alpha\in\mathbb{R}\right\rbrace$. It results that:

\begin{lemma}
\label{lem:equivalence-Pk}
For any $r\ge 1$ and any $\alpha$-free diagrams $D_1,D_2 : r \to n$, 
\begin{align*}
\left(\forall\alpha\in\mathbb{R},~~ \interp{
\InputIfFileExists{r-gn-alpha-to-D_1.tikz}{}{\input{./figures/r-gn-alpha-to-D_1.tikz}}
} = \interp{
\InputIfFileExists{r-gn-alpha-to-D_2.tikz}{}{\input{./figures/r-gn-alpha-to-D_2.tikz}}
}\right) \Leftrightarrow \interp{
\InputIfFileExists{P-to-D1.tikz}{}{\input{./figures/P-to-D1.tikz}}
} = \interp{
\InputIfFileExists{P-to-D2.tikz}{}{\input{./figures/P-to-D2.tikz}}
}
\end{align*}
\end{lemma}
\def\fig{sketch-proof-provability-theorem}
Hence, ~~$\forall\alpha\in\mathbb{R},~\interp{\input{./figures/\fig/\fig_00.tikz}}=\interp{\input{./figures/\fig/\fig_01.tikz}}~~\iff~~\interp{\input{./figures/\fig/\fig_04.tikz}}=\interp{\input{./figures/\fig/\fig_05.tikz}}$. Since $D_1'$, $D_2'$ and $P_r$ are $\dzx_{\pi}$-diagrams, the equality of diagrams on the right hand-side is provable: ~~$\interp{\input{./figures/\fig/\fig_04.tikz}}=\interp{\input{./figures/\fig/\fig_05.tikz}}~~\iff~~\input{./figures/\fig/\fig_04.tikz}\eq{}\input{./figures/\fig/\fig_05.tikz}$

All we need now to finish adapting the result on linear diagrams is the following property:
\[
\InputIfFileExists{n-gn-alpha-to-M_n.tikz}{}{\input{./figures/n-gn-alpha-to-M_n.tikz}}
\]
This property is naturally obtained by induction in the general case provided it is true for the base cases $r\in\{1,2\}$. When $r=1$, the result is obvious. When $r=2$, it deduces easily from:
\[
\InputIfFileExists{beyond-TH-axiom.tikz}{}{\input{./figures/beyond-TH-axiom.tikz}}
\]
In this case:
\def\fig{sketch-proof-provability-theorem}
\[\input{./figures/\fig/\fig_04.tikz}\eq{}\input{./figures/\fig/\fig_05.tikz} ~~\implies~~ \input{./figures/\fig/\fig_02.tikz}\eq{}\input{./figures/\fig/\fig_03.tikz}\]
so finally: ~~$\forall\alpha\in\mathbb{R},~\interp{\input{./figures/\fig/\fig_00.tikz}}=\interp{\input{./figures/\fig/\fig_01.tikz}}~~\iff~~\input{./figures/\fig/\fig_00.tikz}\eq{}\input{./figures/\fig/\fig_01.tikz}$

The result extends naturally to the multi-variable-case.

In conclusion, provided we have the two equations (K) and (P) as axioms, we can prove any sound equation on linear diagrams with constants in $\{0,\pi\}$. With $\dzx^{\text{K,P}}$ denoting the axiomatisation \dzx augmented with the rules (K) and (P):
\begin{theorem}
\label{thm:provability}  
For any \dzx-diagrams $D_1(\vec \alpha)$ and $D_2(\vec \alpha)$ linear in $\vec \alpha=\alpha_1, \ldots, \alpha_k$ with constants in $\{0,\pi\}$:
$$  \forall \vec \alpha \in \mathbb R^k, \interp{D_1(\vec \alpha)}= \interp{D_2(\vec \alpha)}       \iff  \forall \vec \alpha \in \mathbb R^k, \dzx^{\textnormal{K,P}}\vdash D_1(\vec \alpha) = D_2(\vec \alpha)       $$
\end{theorem}

We give two corollaries of this theorem:

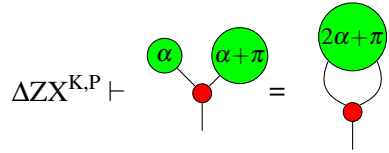
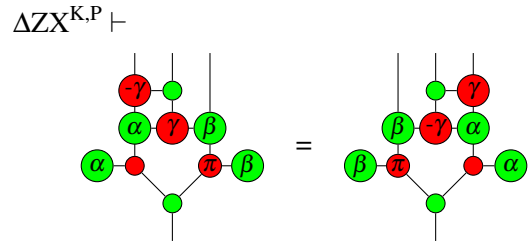
\begin{multicols}{2}
\begin{corollary}
\label{cor:supplementarity}
\[\dzx^{\textnormal{K,P}}\vdash~~
\InputIfFileExists{supplementarity.tikz}{}{\input{./figures/supplementarity.tikz}}
\]
\end{corollary}
\vfill
\begin{corollary}
\label{cor:axiom-C}
\begin{align*}
\dzx^{\textnormal{K,P}}\vdash~~\\
\tag*{
\InputIfFileExists{commutation-of-controls-general-simplified.tikz}{}{\input{./figures/commutation-of-controls-general-simplified.tikz}}
}
\end{align*}
\end{corollary}
\end{multicols}

\section{Axiomatisations for Clifford+T and the General $\dzx$}
\label{sec:clifford+t-triangles}

With the addition of the rules (K) and (P), and the result on linear diagrams (Theorem \ref{thm:provability}), we feel like we are not far from having a complete axiomatisation for the \frag4 of the \dzx-Calculus. However:
\begin{lemma}[\cite{gen-supp}]
$\dzx^{\textnormal{K,P}}\nvdash~~
\InputIfFileExists{bicolor-pi-4-eq-empty.tikz}{}{\input{./figures/bicolor-pi-4-eq-empty.tikz}}
$
\end{lemma}
The language here is different than in \cite{gen-supp}, but the argument is easily adaptable.

\begin{figure*}[!htb]
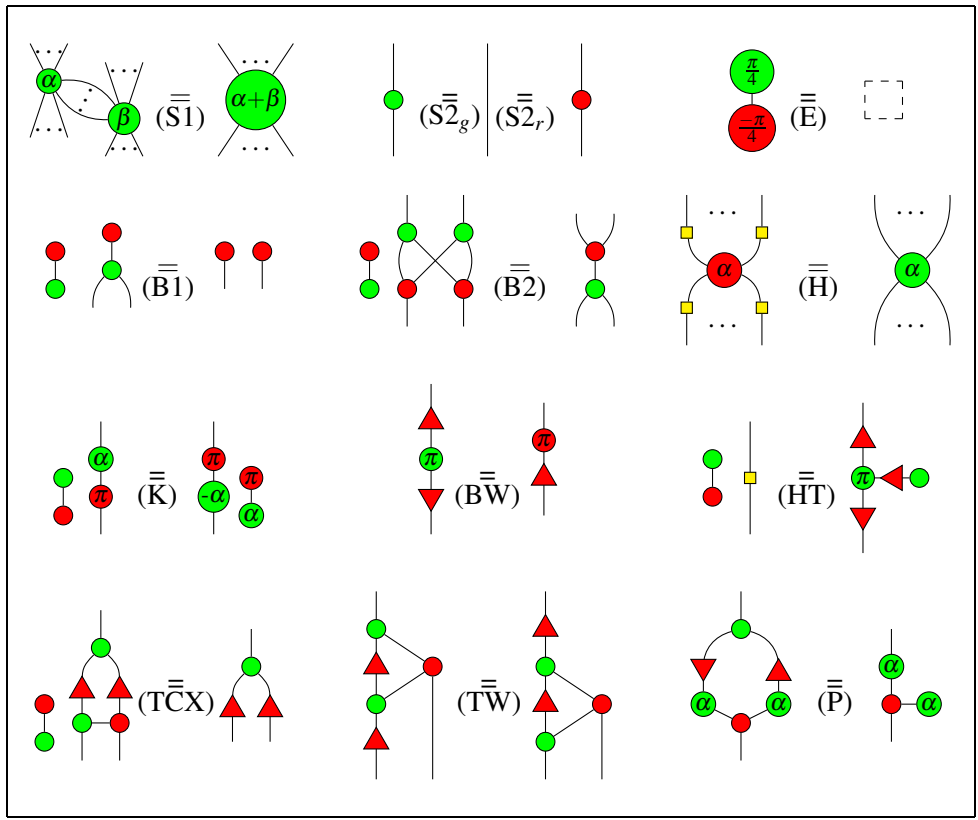

 \centering
 \hypertarget{r:clifford-t-toffoli}{}
 \begin{tabular}{|c@{$\qquad$}c@{$\qquad$}c|}
   \hline
   && \\
   
\InputIfFileExists{spider-1.tikz}{}{\input{./figures/spider-1.tikz}}
&
\InputIfFileExists{s2-green-red.tikz}{}{\input{./figures/s2-green-red.tikz}}
&
\InputIfFileExists{bicolor-pi-4-eq-empty.tikz}{}{\input{./figures/bicolor-pi-4-eq-empty.tikz}}
\\
   && \\
   
\InputIfFileExists{b1s.tikz}{}{\input{./figures/b1s.tikz}}
& 
\InputIfFileExists{b2s.tikz}{}{\input{./figures/b2s.tikz}}
&
\InputIfFileExists{h2.tikz}{}{\input{./figures/h2.tikz}}
\\
   && \\
   
\InputIfFileExists{K-2.tikz}{}{\input{./figures/K-2.tikz}}
&
\InputIfFileExists{BW.tikz}{}{\input{./figures/BW.tikz}}
&
\InputIfFileExists{triangle-hadamard.tikz}{}{\input{./figures/triangle-hadamard.tikz}}
\\
   && \\
   
\InputIfFileExists{cnot-on-triangle-fork.tikz}{}{\input{./figures/cnot-on-triangle-fork.tikz}}
&
\InputIfFileExists{triangle-through-W.tikz}{}{\input{./figures/triangle-through-W.tikz}}
&
\InputIfFileExists{beyond-TH-axiom.tikz}{}{\input{./figures/beyond-TH-axiom.tikz}}
\\
   && \\
   \hline
  \end{tabular}
 \caption[]{Set of rules $\dzx^{\text{K,P,E}}$. The right-hand side of (E) is an empty diagram. (...) denote zero or more wires, while (\protect\rotatebox{45}{\raisebox{-0.4em}{$\cdots$}}) denote one or more wires.
 }
 \label{fig:clifford-t-toffoli-rules}
\end{figure*}

We give a larger axiomatisation denoted $\dzx^{\text{K,P,E}}$ in Figure \ref{fig:clifford-t-toffoli-rules}.
This axiomatisation condenses the initial one ($\dzx$) and the rules (K) and (P), but also replaces (IV) by (E) and gets rid of (Z), though it is easy to show that the new axiomatisation proves the initial one:
\begin{proposition}
\label{prop:TH-derivable-from-clifford-t-toffoli}
$\dzx^{\textnormal{K,P,E}}\vdash\dzx$.
\end{proposition}

As a consequence, this axiomatisation is complete for the $\pi$-fragment of the \dzx-Calculus. It turns out, it is also complete for the \frag4 of the \dzx-Calculus:
\begin{theorem}
\label{thm:clifford-t-toffoli-complete}
The set of rules $\dzx^{\textnormal{K,P,E}}_{\frac{\pi}{4}}$ makes the \frag4 of the \dzx-Calculus complete. For any $D_1$ and $D_2$ diagrams of this fragment:
\[\interp{D_1}=\interp{D_2}\iff\dzx^{\textnormal{K,P,E}}_{\frac{\pi}{4}}\vdash D_1=D_2\]
\end{theorem}

In contrast to the axiomatisations for Clifford+T in \cite{JPV,NgWang-clifford+t}, $\dzx^{\textnormal{K,P,E}}$ got rid of the axioms that were specific to particular angles outside $\{0,\pi\}$, except for \e, which is necessary.

We end up with an axiomatisation for an augmented ZX-Calculus that is complete for the \frag4, as in \cite{NgWang-clifford+t}. However, the method used to achieve completeness is different: in \cite{NgWang-clifford+t}, the authors start from an axiomatisation which is complete in general, and adapt it to the \frag4, whereas in this paper, we began with the easiest fragment of \dzx, and gradually built upon it. As a result, the axiomatisation $\dzx^{\textnormal{K,P,E}}$ uses fewer axioms (14 against $\sim$30) and one fewer generator.

The completeness result on linear diagrams with constants in $\{0,\pi\}$ (Theorem \ref{thm:provability}) can be seen as a generalisation of the completeness of $\dzx_{\pi}$ (Theorem \ref{thm:complete}). Similarly, the completeness result for $\dzx^{\text{K,P,E}}_{\frac{\pi}{4}}$ (Theorem \ref{thm:clifford-t-toffoli-complete}) can be generalised to linear diagrams:
\begin{theorem}
\label{thm:provability-4}  
For any \dzx-diagrams $D_1(\vec \alpha)$ and $D_2(\vec \alpha)$ linear in $\vec \alpha=\alpha_1, \ldots, \alpha_k$ with constants in $\frac{\pi}{4}\mathbb{Z}$:
$$  \forall \vec \alpha \in \mathbb R^k, \interp{D_1(\vec \alpha)}= \interp{D_2(\vec \alpha)}       \iff  \forall \vec \alpha \in \mathbb R^k, \dzx^{\textnormal{K,P,E}}\vdash D_1(\vec \alpha) = D_2(\vec \alpha)       $$
\end{theorem}

\begin{figure}
\centering\boxed{~
\InputIfFileExists{full-completeness-axiom.tikz}{}{\input{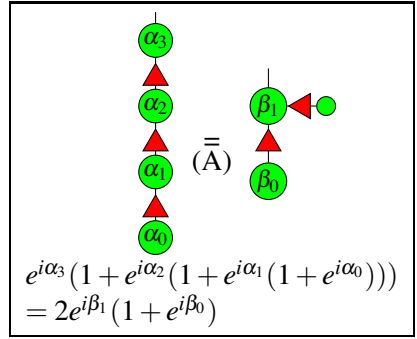}}
~}
 \caption[]{Rule for the completeness of the general $\dzx$-Calculus
 }
 \label{fig:axiom-full}
\end{figure}

In \cite{JPV-universal}, it has been shown that in the ``classical'' ZX-Calculus, completeness from Clifford+T to the general language could be obtained by adding only one rule. Adding precisely the same rule in $\dzx$ would produce the same result. However, we prefer to give a version of the axiom which is more relevant to the current language, in Figure \ref{fig:axiom-full}.

Notice that here again, the axiom is not angle-specific, and that it is non-linear. This is enough to have the completeness:

\begin{theorem}
\label{thm:general-complete}
The set of rules $\dzx^{\textnormal{A,K,P,E}}$ makes the general \dzx-Calculus complete. For any diagrams $D_1$ and $D_2$:
\[\interp{D_1}=\interp{D_2}\iff\dzx^{\textnormal{A,K,P,E}}\vdash D_1=D_2\]
\end{theorem}

\section*{Acknowledgement}
The author acknowledges support from the projects ANR-17-CE25-0009 SoftQPro, ANR-17-CE24-0035 VanQuTe, PIA-GDN/Quantex, and STIC-AmSud 16-STIC-05 FoQCoSS. All diagrams were written with the help of TikZit. The author would like to thank Emmanuel Jeandel and Simon Perdrix for fruitful discussions and suggestions.


\appendix
\section{Appendix}

\subsection{Lemmas}

\begin{multicols}{3}
\begin{lemma}
\label{lem:2-is-sqrt-2-squared}
\[
\InputIfFileExists{2-is-sqrt-2-squared.tikz}{}{\input{./figures/2-is-sqrt-2-squared.tikz}}
\]
\end{lemma}

\begin{lemma}
\label{lem:hopf}
\[
\InputIfFileExists{hopf.tikz}{}{\input{./figures/hopf.tikz}}
\]
\end{lemma}

\begin{lemma}
\label{lem:hadamard-involution}
\[
\InputIfFileExists{hadamard-involution.tikz}{}{\input{./figures/hadamard-involution.tikz}}
\]
\end{lemma}

\begin{lemma}
\label{lem:bicolor-0-pi}
\def\fig{bicolor-0-pi-proof}
\begin{align*}
\input{./figures/\fig/\fig_06.tikz}
\eq{}\input{./figures/\fig/\fig_00.tikz}
\end{align*}
\end{lemma}

\begin{lemma}
\label{lem:ket-0-on-upside-down-triangle}
\def\fig{ket-0-on-upside-down-triangle}
\[\input{./figures/\fig/\fig_00.tikz}\eq{}\input{./figures/\fig/\fig_06.tikz}\]
\end{lemma}

\begin{lemma}
\label{lem:parallel-triangles}
\def\fig{parallel-triangles}
\[\input{./figures/\fig/\fig_00.tikz}\eq{}\input{./figures/\fig/\fig_04.tikz}\]
\end{lemma}

\begin{lemma}
\label{lem:not-triangle-is-symmetrical}
\def\fig{not-triangle-is-symmetrical}
\[\input{./figures/\fig/\fig_00.tikz}\eq{}\input{./figures/\fig/\fig_03.tikz}\]
\end{lemma}

\begin{lemma}
\label{lem:ket-1-on-triangle}
\def\fig{ket-1-on-triangle}
\[\input{./figures/\fig/\fig_00.tikz}\eq{}\input{./figures/\fig/\fig_03.tikz}\]
\end{lemma}

\begin{lemma}
\label{lem:h-loop}
\def\fig{h-loop-from-v2}
\[\input{./figures/\fig/\fig_00.tikz}\eq{}\input{./figures/\fig/\fig_07.tikz}\]
\end{lemma}

\begin{lemma}
\label{lem:k1}
\[
\InputIfFileExists{k1.tikz}{}{\input{./figures/k1.tikz}}
\]
\end{lemma}

\begin{lemma}
\label{lem:multiplying-global-phases}
\[
\InputIfFileExists{multiplying-global-phases.tikz}{}{\input{./figures/multiplying-global-phases.tikz}}
\]
\end{lemma}

\begin{lemma}
\label{lem:control-pi-and-anti-CNOT-commute}
\[
\InputIfFileExists{control-pi-and-anti-CNOT-commute.tikz}{}{\input{./figures/control-pi-and-anti-CNOT-commute.tikz}}
\]
\end{lemma}

\begin{lemma}
\label{lem:ket-1-on-upside-down-triangle}
\def\fig{ket-1-on-upside-down-triangle}
\[\input{./figures/\fig/\fig_00.tikz}\eq{}\input{./figures/\fig/\fig_03.tikz}\]
\end{lemma}

\begin{lemma}
\label{lem:ket-minus-on-upside-down-triangle}
\def\fig{ket-minus-on-upside-down-triangle}
\[\input{./figures/\fig/\fig_00.tikz}\eq{}\input{./figures/\fig/\fig_03.tikz}\]
\end{lemma}

\begin{lemma}
\label{lem:looped-triangle}
\def\fig{looped-triangle}
\[\input{./figures/\fig/\fig_00.tikz}\eq{}\input{./figures/\fig/\fig_05.tikz}\]
\end{lemma}

\begin{lemma}
\label{lem:W-swappable-outputs}
\def\fig{W-swappable-outputs}
\[\input{./figures/\fig/\fig_00.tikz}\eq{}\input{./figures/\fig/\fig_02.tikz}\]
\end{lemma}

\begin{lemma}
\label{lem:triangle-trace}
\def\fig{triangle-trace}
\[\input{./figures/\fig/\fig_00.tikz}\eq{}\input{./figures/\fig/\fig_03.tikz}\]
\end{lemma}

\begin{lemma}
\label{lem:inverse-triangle}
\def\fig{inverse-of-triangle}
\[\input{./figures/\fig/\fig_00.tikz}\eq{}\input{./figures/\fig/\fig_03.tikz}\eq{}\input{./figures/\fig/\fig_07.tikz}\]
\end{lemma}

\begin{lemma}
\label{lem:triangle-hadamard-2}
\def\fig{triangle-hadamard-2}
\[\input{./figures/\fig/\fig_00.tikz}\eq{}\input{./figures/\fig/\fig_04.tikz}\]
\end{lemma}

\begin{lemma}
\label{lem:ctrl-2-and-anti-ctrl-2}
\def\fig{ctrl-2-and-anti-ctrl-2}
\[\input{./figures/\fig/\fig_00.tikz}\eq{}\input{./figures/\fig/\fig_04.tikz}\]
\end{lemma}

\begin{lemma}
\label{lem:parallel-triangle-hadamard}
\def\fig{parallel-triangle-hadamard}
\[\input{./figures/\fig/\fig_00.tikz}\eq{}\input{./figures/\fig/\fig_06.tikz}\]
\end{lemma}

\begin{lemma}
\label{lem:upside-down-triangle-on-W}
\def\fig{upside-down-triangle-on-W-4}
\[\input{./figures/\fig/\fig_06.tikz}\eq{}\input{./figures/\fig/\fig_00.tikz}\]
\end{lemma}

\begin{lemma}
\label{lem:cnot-on-upside-down-triangle-fork}
\def\fig{cnot-on-upside-down-triangle-fork}
\[\input{./figures/\fig/\fig_00.tikz}\eq{}\input{./figures/\fig/\fig_05.tikz}\]
\end{lemma}

\begin{lemma}
\label{lem:transistor-ket-1}
\def\fig{transistor-ket-1}
\[\input{./figures/\fig/\fig_00.tikz}\eq{}\input{./figures/\fig/\fig_03.tikz}\]
\end{lemma}

\end{multicols}

\subsection{Proof of Lemmas}

\begin{proof}[Lemma \ref{lem:2-is-sqrt-2-squared}]
\def\fig{2-is-sqrt-2-squared-proof}
\begin{align*}
\input{./figures/\fig/\fig_00.tikz}
\eq{\so\\\st}\input{./figures/\fig/\fig_01.tikz}
\eq{\bo}\input{./figures/\fig/\fig_02.tikz}
\eq{\so}\input{./figures/\fig/\fig_03.tikz}
\eq{\h}\input{./figures/\fig/\fig_04.tikz}
\end{align*}
\end{proof}

\begin{proof}[Lemma \ref{lem:hopf}]
\def\fig{hopf-proof}
\begin{align*}
\input{./figures/\fig/\fig_00.tikz}
\eq{\st}\input{./figures/\fig/\fig_01.tikz}
\eq{\so}\input{./figures/\fig/\fig_02.tikz}
\eq{\bt}\input{./figures/\fig/\fig_03.tikz}
\eq{\bo}\input{./figures/\fig/\fig_04.tikz}
\eq{\so\\\st}\input{./figures/\fig/\fig_05.tikz}
\end{align*}
\end{proof}

\begin{proof}[Lemma \ref{lem:hadamard-involution}]
\def\fig{hadamard-involution-proof}
\begin{align*}
\input{./figures/\fig/\fig_00.tikz}
\eq{\st}\input{./figures/\fig/\fig_01.tikz}
\eq{\h}\input{./figures/\fig/\fig_02.tikz}
\eq{\st}\input{./figures/\fig/\fig_03.tikz}
\end{align*}
\end{proof}

\begin{proof}[Lemma \ref{lem:bicolor-0-pi}]
\def\fig{bicolor-0-pi-proof}
\begin{align*}
\input{./figures/\fig/\fig_00.tikz}
\eq{\iv\\\h}\input{./figures/\fig/\fig_01.tikz}
\eq{\htt}\input{./figures/\fig/\fig_02.tikz}
\eq{\tz\\\so}\input{./figures/\fig/\fig_03.tikz}
\eq{\iv\\\bo}\input{./figures/\fig/\fig_04.tikz}
\eq{\tz\\\ref{lem:2-is-sqrt-2-squared}\\\iv}\input{./figures/\fig/\fig_05.tikz}
\eq{\iv}\input{./figures/\fig/\fig_06.tikz}
\end{align*}
\end{proof}

\begin{proof}[Lemma \ref{lem:ket-0-on-upside-down-triangle}]
\def\fig{ket-0-on-upside-down-triangle}
\begin{align*}
\input{./figures/\fig/\fig_00.tikz}
\eq{\iv}\input{./figures/\fig/\fig_01.tikz}
\eq{\tz}\input{./figures/\fig/\fig_02.tikz}
\eq{\ref{lem:bicolor-0-pi}\\\bo}\input{./figures/\fig/\fig_03.tikz}
\eq{\tz}\input{./figures/\fig/\fig_04.tikz}
\eq{\htt}\input{./figures/\fig/\fig_05.tikz}
\eq{\h}\input{./figures/\fig/\fig_06.tikz}
\end{align*}
\end{proof}

\begin{proof}[Lemma \ref{lem:parallel-triangles}]
\def\fig{parallel-triangles}
\begin{align*}
\input{./figures/\fig/\fig_00.tikz}
\eq{\st\\\so}\input{./figures/\fig/\fig_01.tikz}
\eq{\iv\\\tcx}\input{./figures/\fig/\fig_02.tikz}
\eq{\ref{lem:ket-0-on-upside-down-triangle}}\input{./figures/\fig/\fig_03.tikz}
\eq{\st\\\so}\input{./figures/\fig/\fig_04.tikz}
\end{align*}
\end{proof}

\begin{proof}[Lemma \ref{lem:not-triangle-is-symmetrical}]
\def\fig{not-triangle-is-symmetrical}
\begin{align*}
\input{./figures/\fig/\fig_00.tikz}
\eq{\bw}\input{./figures/\fig/\fig_01.tikz}
\eq{}\input{./figures/\fig/\fig_02.tikz}
\eq{\bw}\input{./figures/\fig/\fig_03.tikz}
\end{align*}
\end{proof}

\begin{proof}[Lemma \ref{lem:ket-1-on-triangle}]
\def\fig{ket-1-on-triangle}
\begin{align*}
\input{./figures/\fig/\fig_00.tikz}
\eq{\so\\\ref{lem:not-triangle-is-symmetrical}}\input{./figures/\fig/\fig_01.tikz}
\eq{\ref{lem:ket-0-on-upside-down-triangle}}\input{./figures/\fig/\fig_02.tikz}
\eq{\bo\\\ref{lem:bicolor-0-pi}}\input{./figures/\fig/\fig_03.tikz}
\end{align*}
\end{proof}

\begin{proof}[Lemma \ref{lem:h-loop}]
\def\fig{h-loop-from-v2}
\begin{align*}
\input{./figures/\fig/\fig_00.tikz}
\eq{\htt}\input{./figures/\fig/\fig_01.tikz}
\eq{\tcx\\\so}\input{./figures/\fig/\fig_02.tikz}
\eq{\ref{lem:hopf}}\input{./figures/\fig/\fig_03.tikz}\\
\eq{\ref{lem:ket-0-on-upside-down-triangle}\\\iv\\\so\\\st}\input{./figures/\fig/\fig_04.tikz}
\eq{\iv\\\ref{lem:ket-1-on-triangle}}\input{./figures/\fig/\fig_05.tikz}
\eq{\htt}\input{./figures/\fig/\fig_06.tikz}
\eq{\h}\input{./figures/\fig/\fig_07.tikz}
\end{align*}
\end{proof}

\begin{proof}[Lemmas \ref{lem:k1}, \ref{lem:control-pi-and-anti-CNOT-commute}]
By completeness of the $\pi$-fragment of the ZX-Calculus.
\end{proof}

\begin{proof}[Lemma \ref{lem:multiplying-global-phases}]
This proof is classical and uses \ref{lem:k1}, \bo and \so. The proof is  the same whatever the values of $\alpha$ and $\beta$ are.
\end{proof}

\begin{proof}[Lemma \ref{lem:ket-1-on-upside-down-triangle}]
\def\fig{ket-1-on-upside-down-triangle}
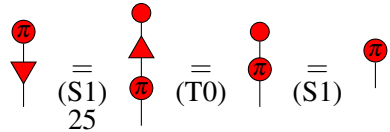
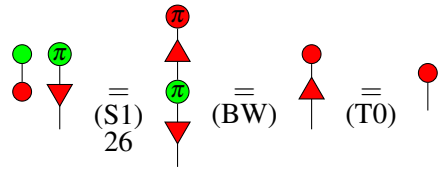
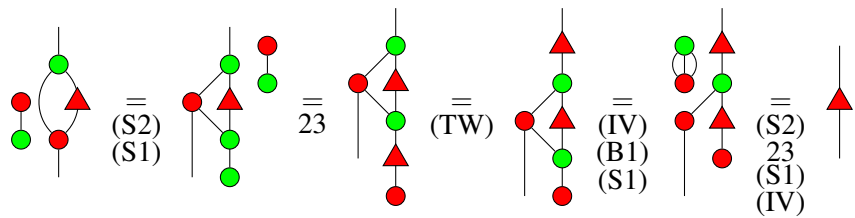
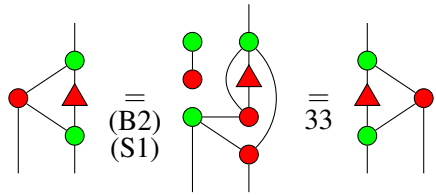
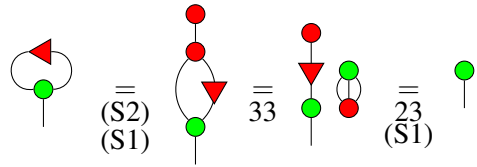
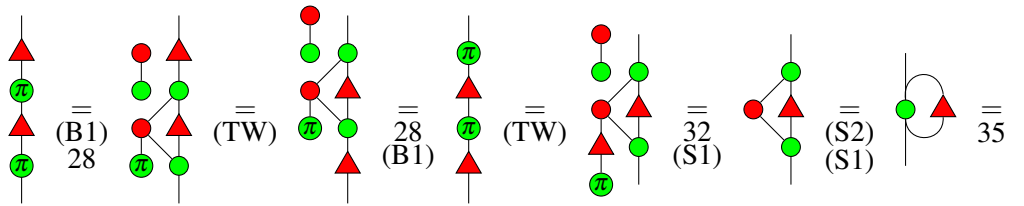
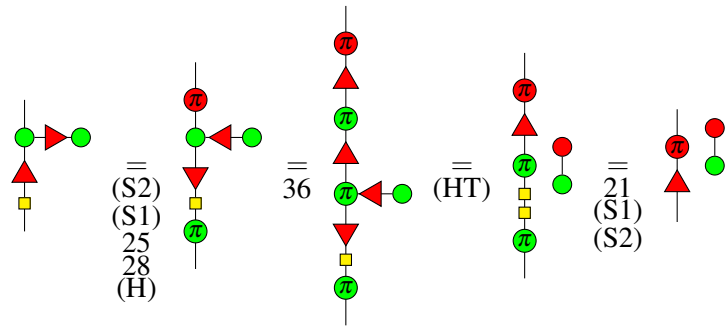
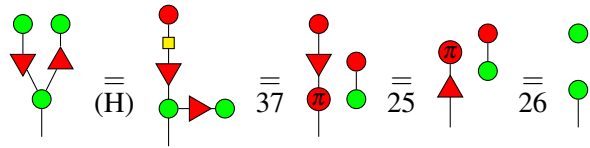
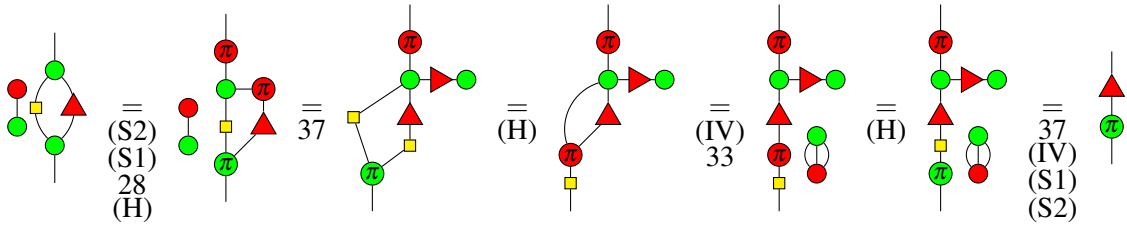
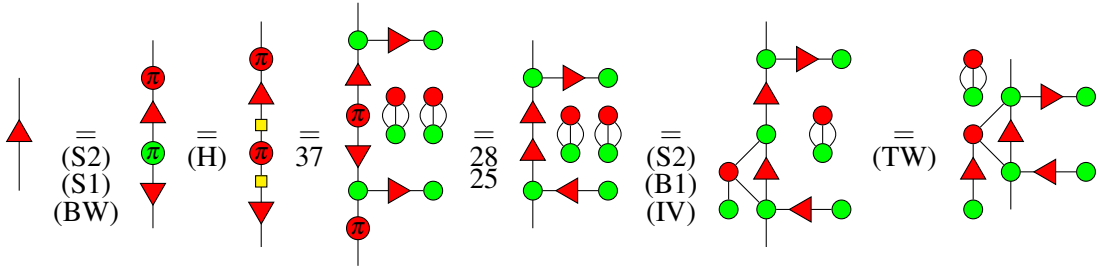
\begin{align*}
\input{./figures/\fig/\fig_00.tikz}
\eq{\so\\\ref{lem:not-triangle-is-symmetrical}}\input{./figures/\fig/\fig_01.tikz}
\eq{\tz}\input{./figures/\fig/\fig_02.tikz}
\eq{\so}\input{./figures/\fig/\fig_03.tikz}
\end{align*}
\end{proof}

\begin{proof}[Lemma \ref{lem:ket-minus-on-upside-down-triangle}]
\def\fig{ket-minus-on-upside-down-triangle}
\begin{align*}
\input{./figures/\fig/\fig_00.tikz}
\eq{\so\\\ref{lem:ket-1-on-triangle}}\input{./figures/\fig/\fig_01.tikz}
\eq{\bw}\input{./figures/\fig/\fig_02.tikz}
\eq{\tz}\input{./figures/\fig/\fig_03.tikz}
\end{align*}
\end{proof}

\begin{proof}[Lemma \ref{lem:looped-triangle}]
\def\fig{looped-triangle}
\begin{align*}
\input{./figures/\fig/\fig_00.tikz}
\eq{\st\\\so}\input{./figures/\fig/\fig_01.tikz}
\eq{\ref{lem:ket-0-on-upside-down-triangle}}\input{./figures/\fig/\fig_02.tikz}
\eq{\tw}\input{./figures/\fig/\fig_03.tikz}
\eq{\iv\\\bo\\\so}\input{./figures/\fig/\fig_04.tikz}
\eq{\st\\\ref{lem:ket-0-on-upside-down-triangle}\\\so\\\iv}\input{./figures/\fig/\fig_05.tikz}
\end{align*}
\end{proof}

\begin{proof}[Lemma \ref{lem:W-swappable-outputs}]
\def\fig{W-swappable-outputs}
\begin{align*}
\input{./figures/\fig/\fig_00.tikz}
\eq{\bt\\\so}\input{./figures/\fig/\fig_01.tikz}
\eq{\ref{lem:looped-triangle}}\input{./figures/\fig/\fig_02.tikz}
\end{align*}
\end{proof}

\begin{proof}[Lemma \ref{lem:triangle-trace}]
\def\fig{triangle-trace}
\begin{align*}
\input{./figures/\fig/\fig_00.tikz}
\eq{\st\\\so}\input{./figures/\fig/\fig_01.tikz}
\eq{\ref{lem:looped-triangle}}\input{./figures/\fig/\fig_02.tikz}
\eq{\ref{lem:ket-0-on-upside-down-triangle}\\\so}\input{./figures/\fig/\fig_03.tikz}
\end{align*}
\end{proof}

\begin{proof}[Lemma \ref{lem:inverse-triangle}]
\def\fig{inverse-of-triangle}
\begin{align*}
\input{./figures/\fig/\fig_00.tikz}
\eq{\bo\\\ref{lem:k1}}\input{./figures/\fig/\fig_01.tikz}
\eq{\tw}\input{./figures/\fig/\fig_02.tikz}
\eq{\ref{lem:k1}\\\bo}\input{./figures/\fig/\fig_03.tikz}
\eq{\tw}\input{./figures/\fig/\fig_04.tikz}
\eq{\ref{lem:ket-minus-on-upside-down-triangle}\\\so}\input{./figures/\fig/\fig_05.tikz}
\eq{\st\\\so}\input{./figures/\fig/\fig_06.tikz}
\eq{\ref{lem:triangle-trace}}\input{./figures/\fig/\fig_07.tikz}
\end{align*}
\end{proof}

\begin{proof}[Lemma \ref{lem:triangle-hadamard-2}]
\def\fig{triangle-hadamard-2}
\begin{align*}
\input{./figures/\fig/\fig_00.tikz}
\eq{\st\\\so\\\ref{lem:not-triangle-is-symmetrical}\\\ref{lem:k1}\\\h}\input{./figures/\fig/\fig_01.tikz}
\eq{\ref{lem:inverse-triangle}}\input{./figures/\fig/\fig_02.tikz}
\eq{\htt}\input{./figures/\fig/\fig_03.tikz}
\eq{\ref{lem:hadamard-involution}\\\so\\\st}\input{./figures/\fig/\fig_04.tikz}
\end{align*}
\end{proof}

\begin{proof}[Lemma \ref{lem:ctrl-2-and-anti-ctrl-2}]
\def\fig{ctrl-2-and-anti-ctrl-2}
\begin{align*}
\input{./figures/\fig/\fig_00.tikz}
\eq{\h}\input{./figures/\fig/\fig_01.tikz}
\eq{\ref{lem:triangle-hadamard-2}}\input{./figures/\fig/\fig_02.tikz}
\eq{\ref{lem:not-triangle-is-symmetrical}}\input{./figures/\fig/\fig_03.tikz}
\eq{\ref{lem:ket-1-on-triangle}}\input{./figures/\fig/\fig_04.tikz}
\end{align*}
\end{proof}

\begin{proof}[Lemma \ref{lem:parallel-triangle-hadamard}]
\def\fig{parallel-triangle-hadamard}
\begin{align*}
\input{./figures/\fig/\fig_00.tikz}
\eq{\st\\\so\\\ref{lem:k1}\\\h}\input{./figures/\fig/\fig_01.tikz}
\eq{\ref{lem:triangle-hadamard-2}}\input{./figures/\fig/\fig_02.tikz}
\eq{\h}\input{./figures/\fig/\fig_03.tikz}
\eq{\iv\\\ref{lem:looped-triangle}}\input{./figures/\fig/\fig_04.tikz}
\eq{\h}\input{./figures/\fig/\fig_05.tikz}
\eq{\ref{lem:triangle-hadamard-2}\\\iv\\\so\\\st}\input{./figures/\fig/\fig_06.tikz}
\end{align*}
\end{proof}

\begin{proof}[Lemma \ref{lem:upside-down-triangle-on-W}]
First:
\def\fig{upside-down-triangle-on-W-1}
\begin{align*}
\input{./figures/\fig/\fig_00.tikz}
\eq{\st\\\so\\\bw}\input{./figures/\fig/\fig_01.tikz}
\eq{\h}\input{./figures/\fig/\fig_02.tikz}
\eq{\ref{lem:triangle-hadamard-2}}\input{./figures/\fig/\fig_03.tikz}
\eq{\ref{lem:k1}\\\ref{lem:not-triangle-is-symmetrical}}\input{./figures/\fig/\fig_04.tikz}
\eq{\st\\\bo\\\iv}\input{./figures/\fig/\fig_05.tikz}
\eq{\tw}\input{./figures/\fig/\fig_06.tikz}
\end{align*}
Moreover, from \ref{lem:triangle-hadamard-2}, we can easily derive:
\begin{align*}
\def\fig{upside-down-triangle-on-W-2}
\input{./figures/\fig/\fig_00.tikz}
\eq{\st\\\so\\\ref{lem:triangle-hadamard-2}}\input{./figures/\fig/\fig_01.tikz}
\eq{\h\\\ref{lem:k1}\\\ref{lem:not-triangle-is-symmetrical}}\input{./figures/\fig/\fig_02.tikz}
\quad\text{and}\quad
\def\fig{upside-down-triangle-on-W-3}
\input{./figures/\fig/\fig_00.tikz}
\eq{\st\\\so\\\ref{lem:not-triangle-is-symmetrical}}\input{./figures/\fig/\fig_01.tikz}
\eq{\ref{lem:triangle-hadamard-2}}\input{./figures/\fig/\fig_02.tikz}
\eq{\h}\input{./figures/\fig/\fig_03.tikz}
\end{align*}
Finally:
\def\fig{upside-down-triangle-on-W-4}
\begin{align*}
\input{./figures/\fig/\fig_00.tikz}
\eq{\h}\input{./figures/\fig/\fig_01.tikz}
\eq{}\input{./figures/\fig/\fig_02.tikz}
\eq{\ref{lem:k1}\\\tw\\\ref{lem:W-swappable-outputs}\\\so}\input{./figures/\fig/\fig_03.tikz}\\
\eq{}\input{./figures/\fig/\fig_04.tikz}
\eq{\bt\\\iv}\input{./figures/\fig/\fig_05.tikz}
\eq{}\input{./figures/\fig/\fig_06.tikz}
\end{align*}
\end{proof}

\begin{proof}[Lemma \ref{lem:cnot-on-upside-down-triangle-fork}]
\def\fig{cnot-on-upside-down-triangle-fork}
\begin{align*}
\input{./figures/\fig/\fig_00.tikz}
\eq{\so\\\ref{lem:not-triangle-is-symmetrical}}\input{./figures/\fig/\fig_01.tikz}
\eq{\ref{lem:k1}\\\ref{lem:not-triangle-is-symmetrical}}\input{./figures/\fig/\fig_02.tikz}
\eq{\ref{lem:k1}}\input{./figures/\fig/\fig_03.tikz}
\eq{\tcx}\input{./figures/\fig/\fig_04.tikz}
\eq{\ref{lem:k1}\\\ref{lem:not-triangle-is-symmetrical}\\\so}\input{./figures/\fig/\fig_05.tikz}
\end{align*}
\end{proof}

\begin{proof}[Lemma \ref{lem:transistor-ket-1}]
\def\fig{transistor-ket-1}
\begin{align*}
\input{./figures/\fig/\fig_00.tikz}
\eq{\so}\input{./figures/\fig/\fig_01.tikz}
\eq{\ref{lem:cnot-on-upside-down-triangle-fork}}\input{./figures/\fig/\fig_02.tikz}
\eq{\tz\\\iv\\\bo\\\ref{lem:ket-0-on-upside-down-triangle}}\input{./figures/\fig/\fig_03.tikz}
\end{align*}
\end{proof}

\subsection{Additional Corollaries of Theorem \ref{thm:provability}}

\begin{corollary}
\label{cor:control-alpha-triangles}
\[\dzx^{\textnormal{K,P}}\vdash~~
\InputIfFileExists{control-alpha-with-triangles.tikz}{}{\input{./figures/control-alpha-with-triangles.tikz}}
\]
\end{corollary}
\begin{corollary}
\label{cor:axiom-C2}
\[\dzx^{\textnormal{K,P}}\vdash~~
\InputIfFileExists{CZ-and-anti-CX-commute.tikz}{}{\input{./figures/CZ-and-anti-CX-commute.tikz}}
\]
\end{corollary}

\subsection{Proofs for Completeness}

\begin{proof}[Proposition \ref{prop:zwcomplete}]
Soundness is obvious, since the rule $iv$ is sound.
  
Now let $D_1$ and $D_2$ be two diagrams of the \zwh-Calculus such that $\interp{D_1}=\interp{D_2}$.We can rewrite $D_1$ and $D_2$ as $D_i = d_i\otimes(\half)^{\otimes   n_i}$ for some integers $n_i$ and diagrams $d_i$ of the \zw-Calculus that do not use the $\half$ symbol. 

We first assume $\interp{D_i}\neq 0$. Notice then that $n_1=n_2\bmod 2$. Indeed $\interp{D_1}=\interp{D_2}\implies \frac{\interp{d_1}}{\sqrt{2}^{n_1}}=\frac{\interp{d_2}}{\sqrt{2}^{n_2}}$. Since $\interp{d_i}$ are matrices over $\mathbb{Z}$, $n_1$ and $n_2$ are either both odd or both even.

First, assume $n_i = 0 \bmod 2$. 
From the new introduced rule, we get that  $\zwh\vdash d_i=D_i\otimes \left(\two\right)^{\otimes \frac{n_i}{2}}$. W.l.o.g. assume $n_1\leq n_2$. Then $\interp{d_1\otimes \left(\two\right)^{\otimes \frac{n_2-n_1}{2}}}=2^{\frac{n_2-n_1}{2}}\interp{d_1}=2^{\frac{n_2}{2}}\interp{D_1} = \interp{d_2}$. Since $d_1\otimes \left(\two\right)^{\otimes \frac{n_2-n_1}{2}}$ and $d_2$ are ZW-diagrams and have the same interpretation, thanks to the completeness of the \zw-Calculus, $\zwh\vdash d_1\otimes \left(\two\right)^{\otimes \frac{n_2-n_1}{2}}=d_2$, which implies $\zwh\vdash d_1\otimes \left(\two\right)^{\otimes \frac{n_2-n_1}{2}}\otimes(\half)^{\otimes n_2}=d_2\otimes(\half)^{\otimes n_2}$ i.e.~$\zwh \vdash D_1 = D_2$.

Now, we can easily show $\zwh\vdash D_1\otimes\half=D_2\otimes\half \iff\zwh \vdash D_1=D_2$, proving the result when $n_i=1\bmod 2$:
\begin{align*}
\zwh\vdash D_1\otimes\half=D_2\otimes\half
\quad\implies\quad \zwh\vdash D_1\otimes\half~\half~\two=D_2\otimes\half~\half~\two\\
\quad\underset{iv}{\implies}\quad \zwh\vdash D_1=D_2
\quad\implies\quad \zwh\vdash D_1\otimes\half=D_2\otimes\half
\end{align*}

Finally, if $\interp{D_1}=\interp{D_2}=0$, then $\interp{d_i}=0$. By completeness of the \zw, $\zw\vdash d_1=d_2$ and $\zw\vdash d_i \otimes \zero = d_i$. Hence, using $iv$ $n_i$ times, $\zwh\vdash d_i = d_i\otimes\zero = d_i\otimes \zero ~(\half)^{\otimes n_i}= d_i\otimes (\half)^{\otimes n_i}=D_i$, so $\zwh\vdash D_1=d_1=d_2=D_2$.
\end{proof}

\begin{proof}[Proposition \ref{prop:ZX-proves-ZW}]
We prove\footnote{The proof is strictly the same as in \cite{JPV}. We give it again here for coherence and to guarantee that all the necessary lemmas have been proven.} here that all the rules of the \zwh-Calculus are preserved by $[.]_X$.\\
$\bullet$ X:
\def\fig{rule-X-proof}
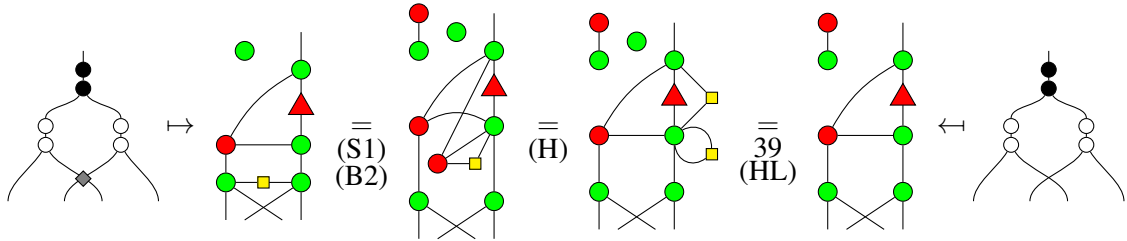
\begin{align*}
\input{./figures/\fig/\fig_00.tikz}\mapsto~~\input{./figures/\fig/\fig_01.tikz}
\eq{\so\\\bt}\input{./figures/\fig/\fig_02.tikz}
\eq{\h}\input{./figures/\fig/\fig_03.tikz}
\eq{\ref{lem:parallel-triangle-hadamard}\\\hl}\input{./figures/\fig/\fig_04.tikz}~~\mapsfrom\input{./figures/\fig/\fig_05.tikz}
\end{align*}
$\bullet$ $0a$, $0c$, $0d$ and $0d'$ come directly from the paradigm \emph{Only Topology Matters}.\\
$\bullet$ $0b$:
\def\fig{rule-0b-proof}
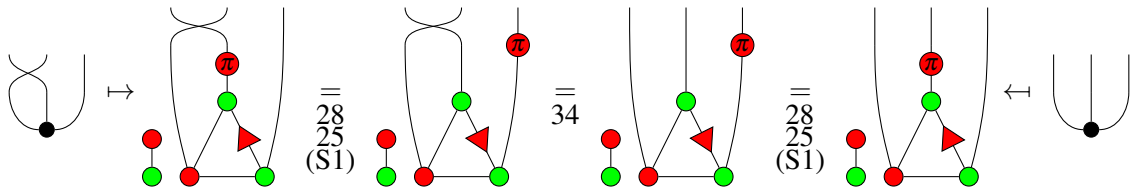
\begin{align*}
\input{./figures/\fig/\fig_00.tikz}~~\mapsto\input{./figures/\fig/\fig_01.tikz}
\eq{\ref{lem:k1}\\\ref{lem:not-triangle-is-symmetrical}\\\so}\input{./figures/\fig/\fig_02.tikz}
\eq{\ref{lem:W-swappable-outputs}}\input{./figures/\fig/\fig_03.tikz}
\eq{\ref{lem:k1}\\\ref{lem:not-triangle-is-symmetrical}\\\so}\input{./figures/\fig/\fig_04.tikz}~\mapsfrom~~\input{./figures/\fig/\fig_05.tikz}
\end{align*}
$\bullet$ $0b'$: Using the result for rule $0b$,
\def\fig{rule-0bp-proof}
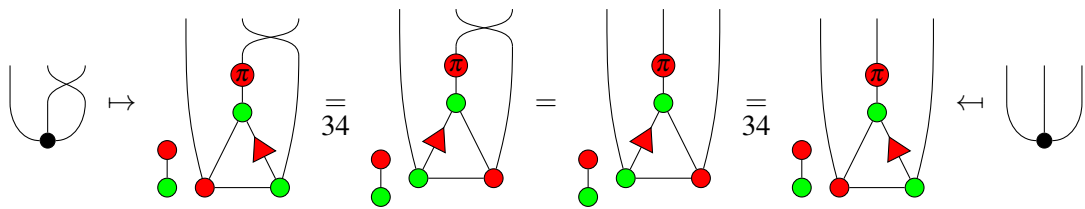
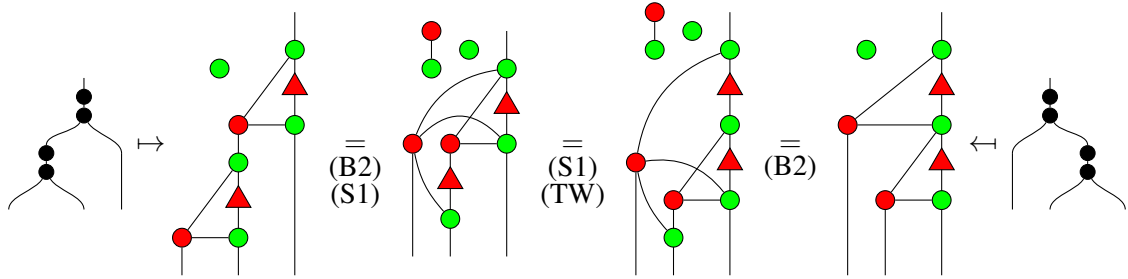
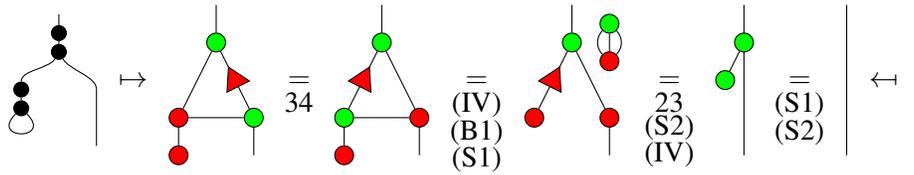
\begin{align*}
\input{./figures/\fig/\fig_00.tikz}~~\mapsto~~\input{./figures/\fig/\fig_01.tikz}
\eq{\ref{lem:W-swappable-outputs}}\input{./figures/\fig/\fig_02.tikz}
\eq{}\input{./figures/\fig/\fig_03.tikz}
\eq{\ref{lem:W-swappable-outputs}}\input{./figures/\fig/\fig_04.tikz}~~\mapsfrom~~\input{./figures/\fig/\fig_05.tikz}
\end{align*}
$\bullet$ $1a$:
\def\fig{rule-1a-proof}
\begin{align*}
\input{./figures/\fig/\fig_00.tikz}~\mapsto\input{./figures/\fig/\fig_01.tikz}
\eq{\bt\\\so}\input{./figures/\fig/\fig_02.tikz}
\eq{\so\\\tw}\input{./figures/\fig/\fig_03.tikz}
\eq{\bt}\input{./figures/\fig/\fig_04.tikz}~\mapsfrom~\input{./figures/\fig/\fig_05.tikz}
\end{align*}
$\bullet$ $1b$:
\def\fig{rule-1b-proof}
\begin{align*}
\input{./figures/\fig/\fig_00.tikz}~~\mapsto~~\input{./figures/\fig/\fig_01.tikz}
\eq{\ref{lem:W-swappable-outputs}}\input{./figures/\fig/\fig_02.tikz}
\eq{\iv\\\bo\\\so}\input{./figures/\fig/\fig_03.tikz}
\eq{\ref{lem:ket-0-on-upside-down-triangle}\\\st\\\iv}\input{./figures/\fig/\fig_04.tikz}
\eq{\so\\\st}\input{./figures/\fig/\fig_05.tikz}~~\mapsfrom~~\input{./figures/\fig/\fig_05.tikz}
\end{align*}
$\bullet$ $1c$, $1d$, $2a$ and $2b$ come directly from the spider
rules \so and \st.\\
$\bullet$ $3a$ is Lemma \ref{lem:k1}.\\
$\bullet$ $3b$:
\def\fig{rule-3b-proof}
\begin{align*}
\input{./figures/\fig/\fig_00.tikz}\quad\mapsto\quad\input{./figures/\fig/\fig_01.tikz}
\eq[\quad]{\ref{lem:k1}\\\so}\input{./figures/\fig/\fig_02.tikz}\quad\mapsfrom\quad\input{./figures/\fig/\fig_03.tikz}
\end{align*}
$\bullet$ $4$ comes from the spider rule \so.\\
$\bullet$ $5a$: We will need a few steps to prove this equality.\\
\step \label{step:one}
\def\fig{anti-control-pi-and-control-triangle-commute}
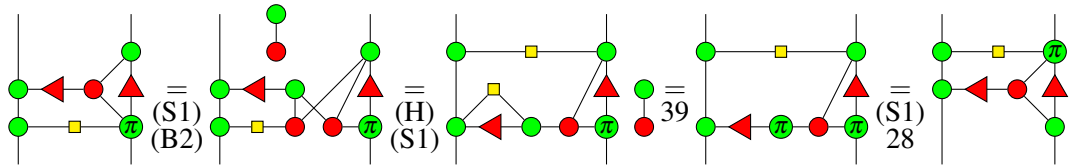
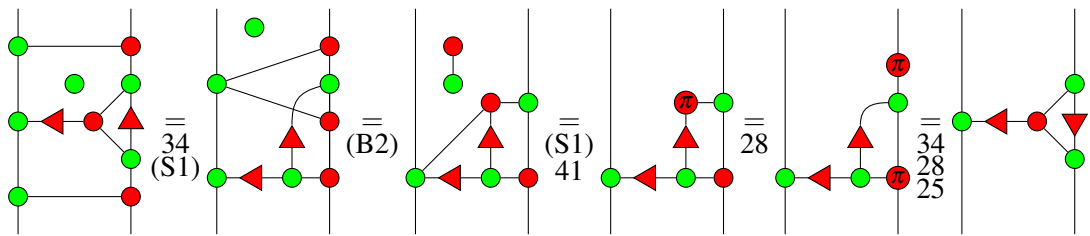
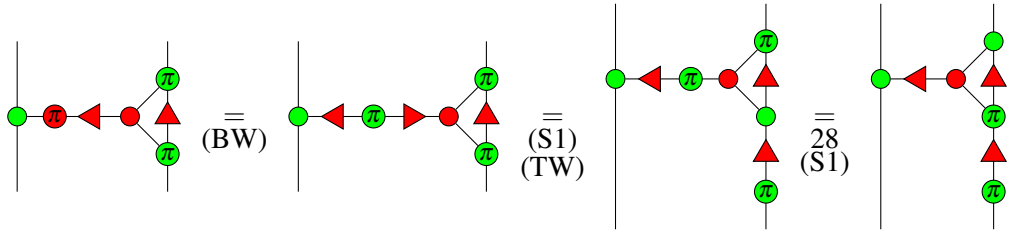
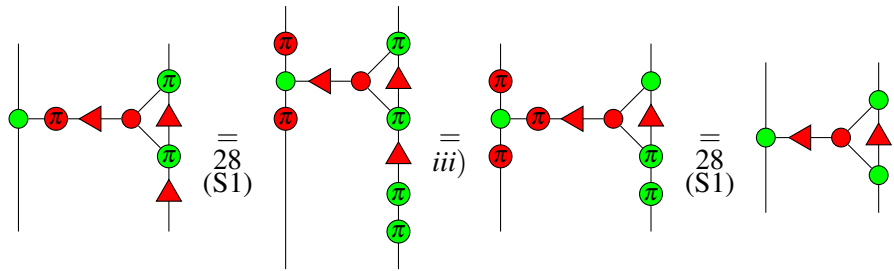
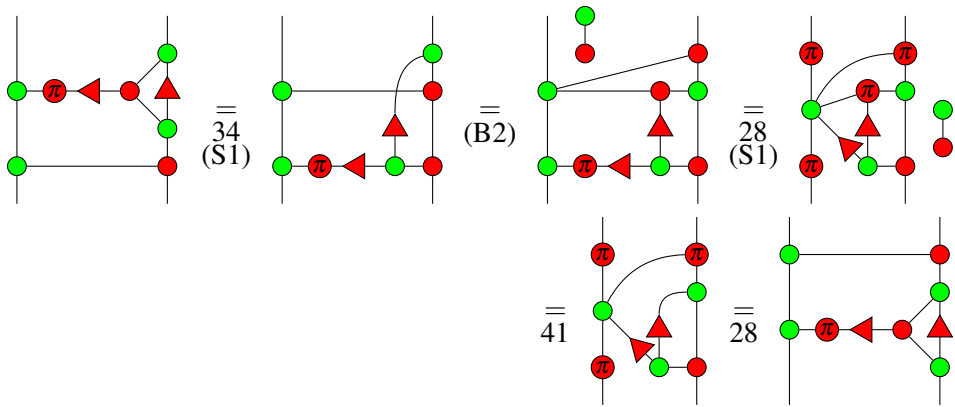
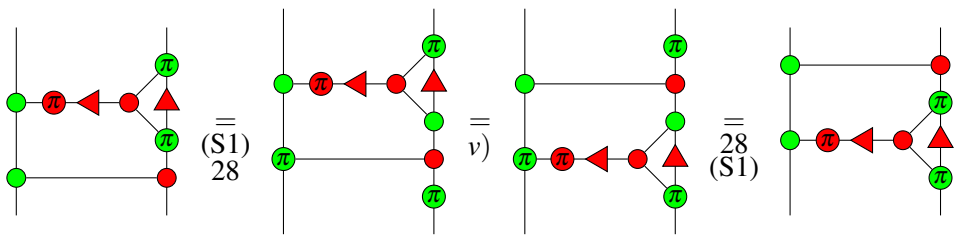
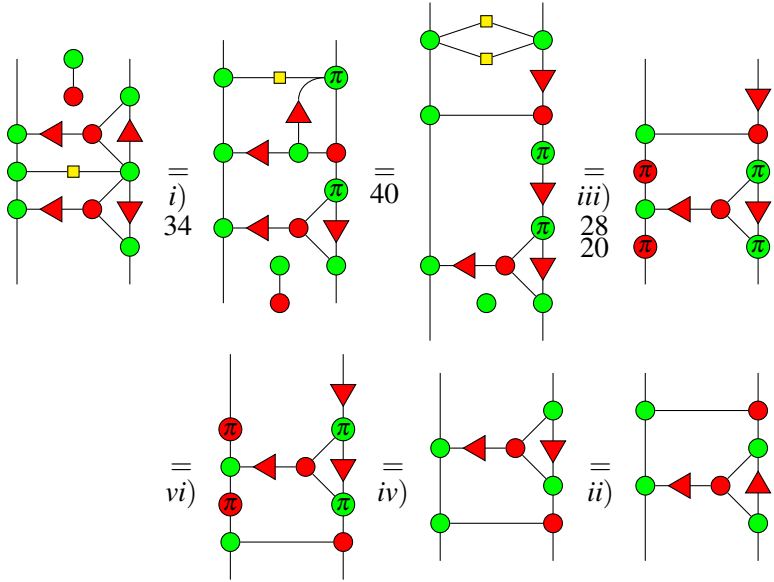
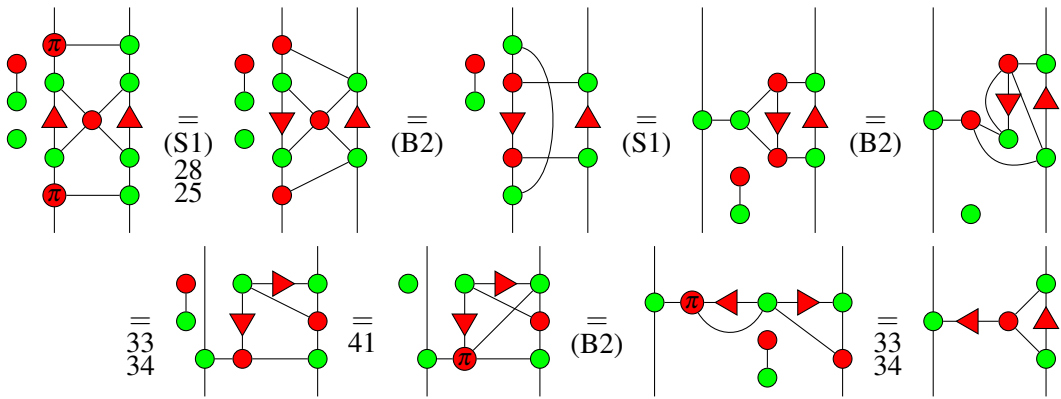
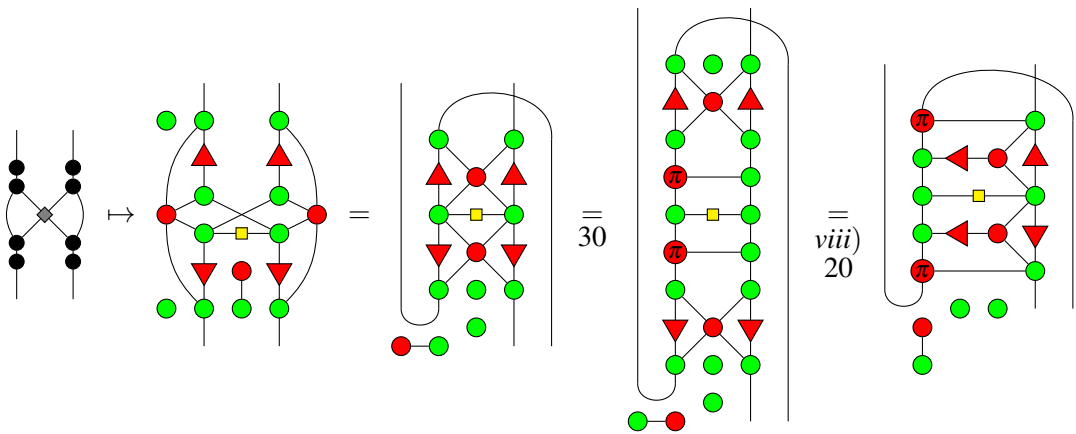
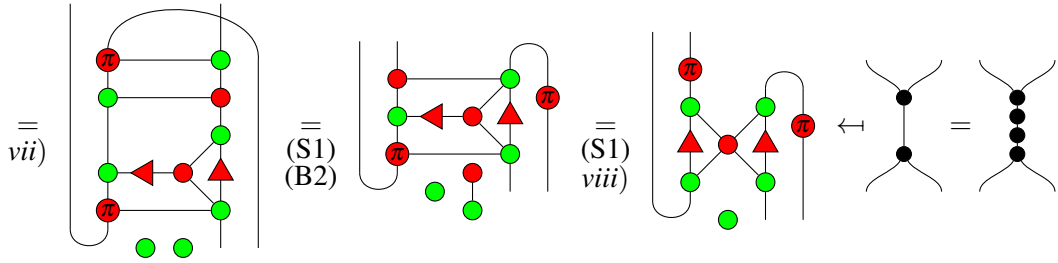
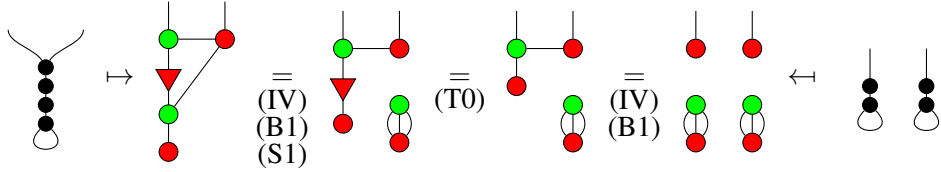
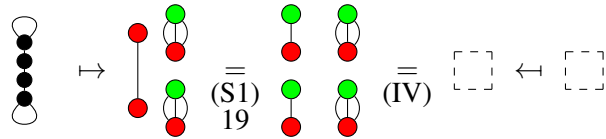
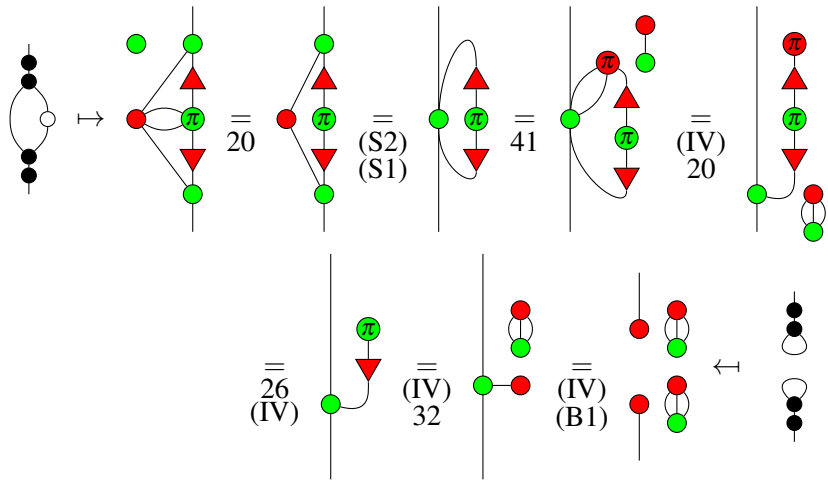
\begin{align*}
\input{./figures/\fig/\fig_00.tikz}
\eq[]{\so\\\bt}\input{./figures/\fig/\fig_01.tikz}
\eq[]{\h\\\so}\input{./figures/\fig/\fig_02.tikz}
\eq[]{\ref{lem:parallel-triangle-hadamard}}\input{./figures/\fig/\fig_03.tikz}
\eq[]{\so\\\ref{lem:k1}}\input{./figures/\fig/\fig_04.tikz}
\end{align*}
\step \label{step:two}
\def\fig{CNOT-control-triangle-CNOT-is-control-transpose-triangle}
\begin{align*}
\input{./figures/\fig/\fig_00.tikz}
\eq[]{\ref{lem:W-swappable-outputs}\\\so}\input{./figures/\fig/\fig_01.tikz}
\eq[]{\bt}\input{./figures/\fig/\fig_02.tikz}
\eq[]{\so\\\ref{lem:cnot-on-upside-down-triangle-fork}}\input{./figures/\fig/\fig_03.tikz}
\eq[]{\ref{lem:k1}}\input{./figures/\fig/\fig_04.tikz}
\eq[]{\ref{lem:W-swappable-outputs}\\\ref{lem:k1}\\\ref{lem:not-triangle-is-symmetrical}}\input{./figures/\fig/\fig_05.tikz}
\end{align*}
\step \label{step:three}
\def\fig{anti-control-inverse-triangle-is-control-triangle-times-inverse-triangle}
\begin{align*}
\input{./figures/\fig/\fig_00.tikz}
\eq{\bw}\input{./figures/\fig/\fig_01.tikz}
\eq{\so\\\tw}\input{./figures/\fig/\fig_02.tikz}
\eq{\ref{lem:k1}\\\so}\input{./figures/\fig/\fig_03.tikz}
\end{align*}
\step \label{step:three-bis}
\def\fig{anti-control-inverse-triangle-is-control-triangle-times-inverse-triangle-bis}
\begin{align*}
\input{./figures/\fig/\fig_00.tikz}
\eq{\ref{lem:k1}\\\so}\input{./figures/\fig/\fig_01.tikz}
\eq{\ref{step:three}}\input{./figures/\fig/\fig_02.tikz}
\eq{\ref{lem:k1}\\\so}\input{./figures/\fig/\fig_03.tikz}
\end{align*}
\step \label{step:four}
\def\fig{anti-control-triangle-and-CNOT-commute}
\begin{align*}
\input{./figures/\fig/\fig_00.tikz}
\eq{\ref{lem:W-swappable-outputs}\\\so}\input{./figures/\fig/\fig_01.tikz}
\eq{\bt}\input{./figures/\fig/\fig_02.tikz}
\eq{\ref{lem:k1}\\\so}\input{./figures/\fig/\fig_03.tikz}\\
\eq{\ref{lem:cnot-on-upside-down-triangle-fork}}\input{./figures/\fig/\fig_04.tikz}
\eq{\ref{lem:k1}}\input{./figures/\fig/\fig_05.tikz}
\end{align*}
\step \label{step:five}
\def\fig{anti-control-inverse-triangle-and-CNOT-commute}
\begin{align*}
\input{./figures/\fig/\fig_00.tikz}
\eq{\so\\\ref{lem:k1}}\input{./figures/\fig/\fig_01.tikz}
\eq{\ref{step:four}}\input{./figures/\fig/\fig_02.tikz}
\eq{\ref{lem:k1}\\\so}\input{./figures/\fig/\fig_03.tikz}
\end{align*}
\step \label{step:six}
\def\fig{control-not-ug-is-symmetrical-proof}
\begin{align*}
\input{./figures/\fig/\fig_00.tikz}
\eq{\ref{step:one}\\\ref{lem:W-swappable-outputs}}\input{./figures/\fig/\fig_01.tikz}
\eq{\ref{lem:upside-down-triangle-on-W}}\input{./figures/\fig/\fig_02.tikz}
\eq{\ref{step:three}\\\ref{lem:k1}\\\ref{lem:hopf}}\input{./figures/\fig/\fig_03.tikz}\\
\eq{\ref{step:five}}\input{./figures/\fig/\fig_04.tikz}
\eq{\ref{step:three-bis}}\input{./figures/\fig/\fig_05.tikz}
\eq{\ref{step:two}}\input{./figures/\fig/\fig_06.tikz}
\end{align*}
\step \label{step:seven}
\def\fig{2-diagrams-of-control-triangle}
\begin{align*}
\input{./figures/\fig/\fig_00.tikz}
\eq{\so\\\ref{lem:k1}\\\ref{lem:not-triangle-is-symmetrical}}\input{./figures/\fig/\fig_01.tikz}
\eq{\bt}\input{./figures/\fig/\fig_02.tikz}
\eq{\so}\input{./figures/\fig/\fig_03.tikz}
\eq{\bt}\input{./figures/\fig/\fig_04.tikz}\\
\eq{\ref{lem:looped-triangle}\\\ref{lem:W-swappable-outputs}}\input{./figures/\fig/\fig_05.tikz}
\eq{\ref{lem:cnot-on-upside-down-triangle-fork}}\input{./figures/\fig/\fig_06.tikz}
\eq{\bt}\input{./figures/\fig/\fig_07.tikz}
\eq{\ref{lem:looped-triangle}\\\ref{lem:W-swappable-outputs}}\input{./figures/\fig/\fig_08.tikz}
\end{align*}
Finally,
\def\fig{rule-5a-proof}
\begin{align*}
\input{./figures/\fig/\fig_00.tikz}~~\mapsto~~\input{./figures/\fig/\fig_01.tikz}
\eq{}\input{./figures/\fig/\fig_02.tikz}
\eq{\ref{lem:control-pi-and-anti-CNOT-commute}}\input{./figures/\fig/\fig_03.tikz}
\eq{\ref{step:seven} \\\ref{lem:hopf}}\input{./figures/\fig/\fig_04.tikz}\\
\eq{\ref{step:six}}\input{./figures/\fig/\fig_05.tikz}
\eq{\so\\\bt}\input{./figures/\fig/\fig_06.tikz}
\eq{\so\\\ref{step:seven}}\input{./figures/\fig/\fig_07.tikz}~~\mapsfrom\!\!\input{./figures/\fig/\fig_08.tikz}=\input{./figures/\fig/\fig_09.tikz}
\end{align*}
$\bullet$ $5b$:
\def\fig{rule-5b-proof}
\begin{align*}
\input{./figures/\fig/\fig_00.tikz}~~\mapsto~~\input{./figures/\fig/\fig_01.tikz}
\eq{\iv\\\bo\\\so}\input{./figures/\fig/\fig_02.tikz}
\eq{\tz}\input{./figures/\fig/\fig_03.tikz}
\eq{\iv\\\bo}\input{./figures/\fig/\fig_04.tikz}~~\mapsfrom~~\input{./figures/\fig/\fig_05.tikz}
\end{align*}
$\bullet$ $5c$:
\def\fig{rule-5c-proof}
\begin{align*}
\input{./figures/\fig/\fig_00.tikz}~~\mapsto~~\input{./figures/\fig/\fig_01.tikz}
\eq{\so\\\ref{lem:2-is-sqrt-2-squared}}\input{./figures/\fig/\fig_02.tikz}
\eq{\iv}\input{./figures/\fig/\fig_03.tikz}~~\mapsfrom~~\input{./figures/\fig/\fig_03.tikz}
\end{align*}
$\bullet$ $5d$:
\def\fig{rule-5d-proof}
\begin{align*}
\input{./figures/\fig/\fig_00.tikz}~~\mapsto~~\input{./figures/\fig/\fig_01.tikz}
\eq{\ref{lem:hopf}}\input{./figures/\fig/\fig_02.tikz}
\eq{\st\\\so}\input{./figures/\fig/\fig_03.tikz}
\eq{\ref{lem:cnot-on-upside-down-triangle-fork}}\input{./figures/\fig/\fig_04.tikz}
\eq{\iv\\\ref{lem:hopf}}\input{./figures/\fig/\fig_05.tikz}\\
\eq{\ref{lem:ket-1-on-triangle}\\\iv}\input{./figures/\fig/\fig_06.tikz}
\eq{\iv\\\ref{lem:ket-minus-on-upside-down-triangle}}\input{./figures/\fig/\fig_07.tikz}
\eq{\iv\\\bo}\input{./figures/\fig/\fig_08.tikz}~~\mapsfrom~~\input{./figures/\fig/\fig_09.tikz}
\end{align*}
$\bullet$ $6a$: Thanks to the rule X we can get rid of \scalebox{0.6}{
\InputIfFileExists{control-pi.tikz}{}{\input{./figures/control-pi.tikz}}
} induced by the crossing. Then,
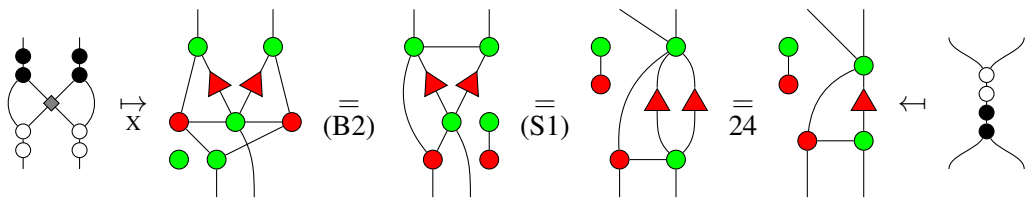
\def\fig{rule-6a-proof}
\begin{align*}
\input{./figures/\fig/\fig_00.tikz}~~\underset{\text{X}}{\mapsto}~~\input{./figures/\fig/\fig_01.tikz}
\eq{\bt}\input{./figures/\fig/\fig_02.tikz}
\eq{\so}\input{./figures/\fig/\fig_03.tikz}
\eq{\ref{lem:parallel-triangles}}\input{./figures/\fig/\fig_04.tikz}~~\mapsfrom~~\input{./figures/\fig/\fig_05.tikz}
\end{align*}
$\bullet$ $6b$ is exactly the copy rule \bo.\\
$\bullet$ $6c$:
\def\fig{rule-6c-proof}
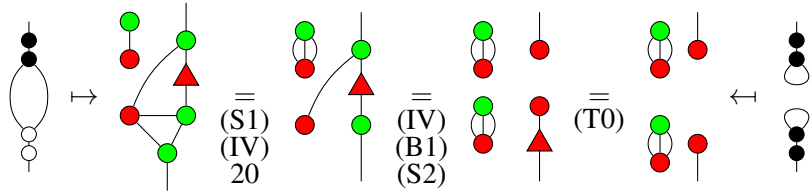
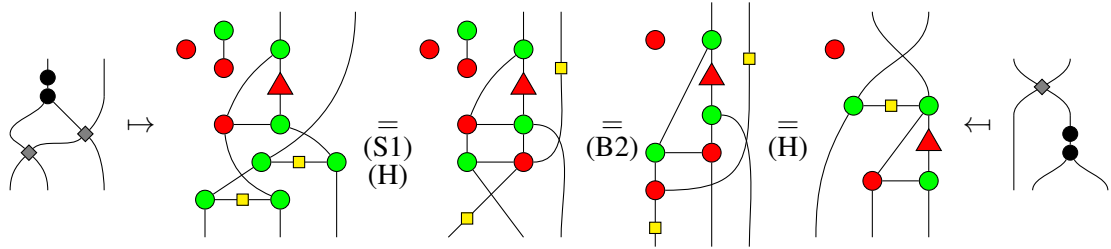
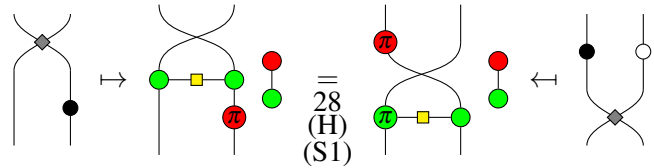
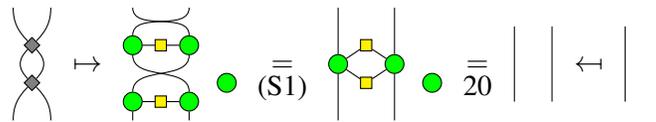
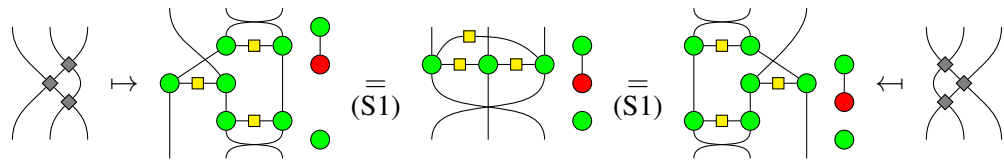
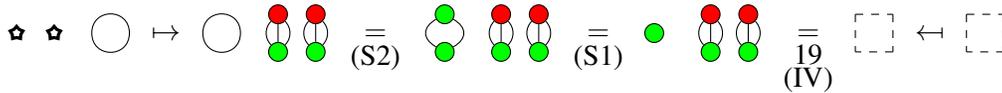
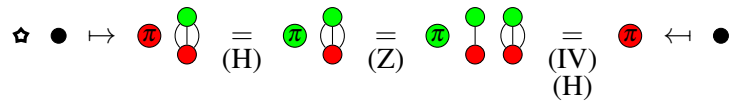
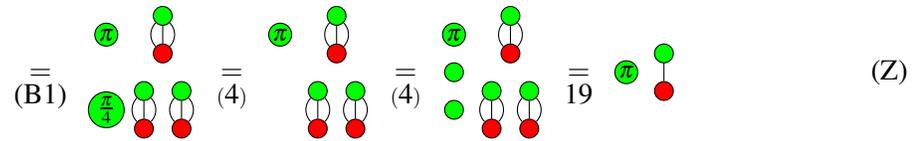
\begin{align*}
\input{./figures/\fig/\fig_00.tikz}~~\mapsto~~\input{./figures/\fig/\fig_01.tikz}
\eq{\so\\\iv\\\ref{lem:hopf}}\input{./figures/\fig/\fig_02.tikz}
\eq{\iv\\\bo\\\st}\input{./figures/\fig/\fig_03.tikz}
\eq{\tz}\input{./figures/\fig/\fig_04.tikz}~~\mapsfrom\input{./figures/\fig/\fig_05.tikz}
\end{align*}
$\bullet$ $7a$:
\def\fig{rule-7a-proof}
\begin{align*}
\input{./figures/\fig/\fig_00.tikz}~~\mapsto~~\input{./figures/\fig/\fig_01.tikz}
\eq[]{\so\\\h}\input{./figures/\fig/\fig_02.tikz}
\eq[]{\bt}\input{./figures/\fig/\fig_03.tikz}
\eq[]{\h}\input{./figures/\fig/\fig_04.tikz}~~\mapsfrom~~\input{./figures/\fig/\fig_05.tikz}
\end{align*}
$\bullet$ $7b$:
\def\fig{rule-7b-proof}
\begin{align*}
\input{./figures/\fig/\fig_00.tikz}~~\mapsto~~\input{./figures/\fig/\fig_01.tikz}
\eq{\ref{lem:k1}\\\h\\\so}\input{./figures/\fig/\fig_02.tikz}
~~\mapsfrom~~\input{./figures/\fig/\fig_04.tikz}
\end{align*}
$\bullet$ R$_2$:
\def\fig{rule-r2-proof}
\begin{align*}
\input{./figures/\fig/\fig_00.tikz}~~\mapsto~~\input{./figures/\fig/\fig_01.tikz}
\eq{\so}\input{./figures/\fig/\fig_02.tikz}
\eq{\ref{lem:hopf}}\input{./figures/\fig/\fig_03.tikz}~~\mapsfrom~~\input{./figures/\fig/\fig_03.tikz}
\end{align*}
$\bullet$ R$_3$:
\def\fig{rule-r3-proof}
\begin{align*}
\input{./figures/\fig/\fig_00.tikz}~~\mapsto~~\input{./figures/\fig/\fig_01.tikz}
\eq{\so}\input{./figures/\fig/\fig_02.tikz}
\eq{\so}\input{./figures/\fig/\fig_03.tikz}~~\mapsfrom~~\input{./figures/\fig/\fig_04.tikz}
\end{align*}
$\bullet$ $iv$: 
\def\fig{rule-half-proof}
\begin{align*}
\input{./figures/\fig/\fig_00.tikz}~~\mapsto~~\input{./figures/\fig/\fig_01.tikz}
\eq{\st}\input{./figures/\fig/\fig_02.tikz}
\eq{\so}\input{./figures/\fig/\fig_03.tikz}
\eq{\ref{lem:2-is-sqrt-2-squared}\\\iv}\input{./figures/\fig/\fig_04.tikz}
~~\mapsfrom~~\input{./figures/\fig/\fig_05.tikz}
\end{align*}
$\bullet$ $z$:
\def\fig{rule-zero-proof}
\begin{align*}
\input{./figures/\fig/\fig_00.tikz}~~\mapsto~~\input{./figures/\fig/\fig_01.tikz}
\eq{\h}\input{./figures/\fig/\fig_02.tikz}
\eq{\z}\input{./figures/\fig/\fig_03.tikz}
\eq{\iv\\\h}\input{./figures/\fig/\fig_04.tikz}
~~\mapsfrom~~\input{./figures/\fig/\fig_05.tikz}
\end{align*}
\end{proof}

\begin{proof}[Proposition \ref{prop:TH-derivable-from-clifford-t-toffoli}]
All the equations in Figure \ref{fig:ZX-rules} are present in the new axiomatisation, except \iv, \z and \tz. Proving \iv and Lemmas \ref{lem:2-is-sqrt-2-squared} and \ref{lem:hopf} from \so, \st, \bo, \bt and \e, is classical \cite{gen-supp}. Notice also that we can easily derive the colour-swapped version of all the rules. Then:
\def\fig{ket-minus-on-upside-down-triangle-from-v2}
\begin{align}
\label{eq:v2-ket-minus-on-upside-down-triangle}
\input{./figures/\fig/\fig_00.tikz}
\eq{\so\\\iv\\\ref{lem:2-is-sqrt-2-squared}}\input{./figures/\fig/\fig_01.tikz}
\eq{\so\\\bo}\input{./figures/\fig/\fig_02.tikz}
\eq{\p}\input{./figures/\fig/\fig_03.tikz}
\eq{\bo}\input{./figures/\fig/\fig_04.tikz}
\eq{\htt}\input{./figures/\fig/\fig_05.tikz}
\eq{\iv}\input{./figures/\fig/\fig_06.tikz}
\end{align}
\def\fig{bicolor-0-pi-from-v2}
\begin{align}
\label{eq:v2-bicolor-0-pi}
\input{./figures/\fig/\fig_00.tikz}
\eq{\iv}\input{./figures/\fig/\fig_01.tikz}
\eq{\so}\input{./figures/\fig/\fig_02.tikz}
\eq{\kt}\input{./figures/\fig/\fig_03.tikz}
\eq{\st}\input{./figures/\fig/\fig_04.tikz}
\eq{\so}\input{./figures/\fig/\fig_05.tikz}
\end{align}
\def\fig{not-triangle-is-symmetrical}
\begin{align}
\label{eq:v2-not-triangle-symmetrical}
\input{./figures/\fig/\fig_00.tikz}
\eq{\bw}\input{./figures/\fig/\fig_01.tikz}
\eq{}\input{./figures/\fig/\fig_02.tikz}
\eq{\bw}\input{./figures/\fig/\fig_03.tikz}
\end{align}
\def\fig{ket-0-on-triangle-from-v2}
\begin{align*}
\label{eq:v2-ket-0-on-triangle}
\input{./figures/\fig/\fig_00.tikz}
\eq{\iv\\\bo\\(\ref{eq:v2-bicolor-0-pi})}\input{./figures/\fig/\fig_01.tikz}
\eq{\st\\\so\\(\ref{eq:v2-not-triangle-symmetrical})\\(\ref{eq:v2-ket-minus-on-upside-down-triangle})}\input{./figures/\fig/\fig_02.tikz}
\eq{\kt\\(\ref{eq:v2-not-triangle-symmetrical})}\input{./figures/\fig/\fig_03.tikz}
\eq{\bw\\\so\\\st\\\kt}\input{./figures/\fig/\fig_04.tikz}
\eq{\so\\\bo\\(\ref{eq:v2-bicolor-0-pi})}\input{./figures/\fig/\fig_05.tikz}
\eq{(\ref{eq:v2-ket-minus-on-upside-down-triangle})}\input{./figures/\fig/\fig_06.tikz}\tag{T0}
\end{align*}
It only remains to prove \z. Notice that it is not used in the proof of Lemmas \ref{lem:h-loop} and \ref{lem:inverse-triangle}, so they can be derived. Also, without \z, 
\InputIfFileExists{bicolor-0-alpha.tikz}{}{\input{./figures/bicolor-0-alpha.tikz}}
 is derivable (see \cite{simplified-stabilizer}), so now:
\def\fig{zero-alpha}
\begin{align}
\label{eq:v2-zero-alpha}
\input{./figures/\fig/\fig_00.tikz}
\eq{\so\\\bo}\input{./figures/\fig/\fig_01.tikz}
\eq{\p}\input{./figures/\fig/\fig_02.tikz}
\eq{\st\\\so\\\ref{lem:inverse-triangle}}\input{./figures/\fig/\fig_03.tikz}
\eq{\ref{lem:hopf}}\input{./figures/\fig/\fig_04.tikz}
\eq{}\input{./figures/\fig/\fig_05.tikz}
\eq{}\input{./figures/\fig/\fig_06.tikz}
\end{align}
\def\fig{zero-rule-proof}
\begin{align*}
\input{./figures/\fig/\fig_00.tikz}
\eq{\e}\input{./figures/\fig/\fig_01.tikz}
\eq{\ref{lem:hopf}}\input{./figures/\fig/\fig_02.tikz}
\eq{\ref{lem:inverse-triangle}}\input{./figures/\fig/\fig_03.tikz}
\eq{\p}\input{./figures/\fig/\fig_04.tikz}\\
\eq{\bo}\input{./figures/\fig/\fig_05.tikz}
\eq{(\ref{eq:v2-zero-alpha})}\input{./figures/\fig/\fig_06.tikz}
\eq{(\ref{eq:v2-zero-alpha})}\input{./figures/\fig/\fig_07.tikz}
\eq{\ref{lem:2-is-sqrt-2-squared}}\input{./figures/\fig/\fig_08.tikz}\tag{Z}
\end{align*}
\end{proof}

\begin{proof}[Theorem \ref{thm:clifford-t-toffoli-complete}]
We remind the rules that make the \frag4 complete \cite{JPV}:\\
\needspace{11em}\noindent\titlerule{~~$\zx_{\pi/4}$~~}
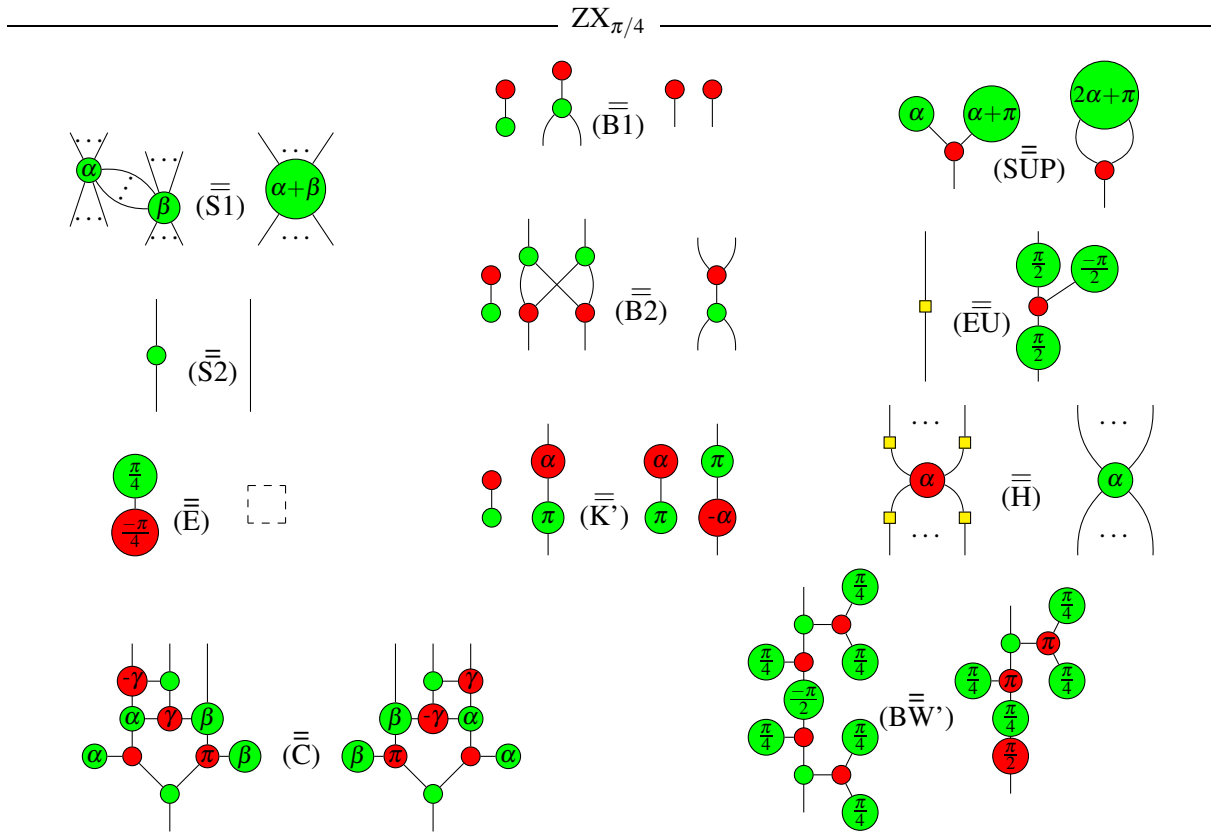
\begin{multicols}{3}
\[
\InputIfFileExists{spider-1.tikz}{}{\input{./figures/spider-1.tikz}}
\]
\[
\InputIfFileExists{s2-simple.tikz}{}{\input{./figures/s2-simple.tikz}}
\]
\[
\InputIfFileExists{bicolor-pi-4-eq-empty.tikz}{}{\input{./figures/bicolor-pi-4-eq-empty.tikz}}
\]
\[
\InputIfFileExists{b1s.tikz}{}{\input{./figures/b1s.tikz}}
\]
\[
\InputIfFileExists{b2s.tikz}{}{\input{./figures/b2s.tikz}}
\]
\[
\InputIfFileExists{k2s.tikz}{}{\input{./figures/k2s.tikz}}
\]
\[
\InputIfFileExists{supplementarity-label.tikz}{}{\input{./figures/supplementarity-label.tikz}}
\]
\[
\InputIfFileExists{euler-decomp-scalar-free.tikz}{}{\input{./figures/euler-decomp-scalar-free.tikz}}
\]
\[
\InputIfFileExists{h2.tikz}{}{\input{./figures/h2.tikz}}
\]
\end{multicols}
\noindent\begin{minipage}{\columnwidth}
\begin{multicols}{2}
\[
\InputIfFileExists{commutation-of-controls-general-simplified-label.tikz}{}{\input{./figures/commutation-of-controls-general-simplified-label.tikz}}
\]
\[
\InputIfFileExists{BW-pure-zx.tikz}{}{\input{./figures/BW-pure-zx.tikz}}
\]
\end{multicols}
\rule{\columnwidth}{0.5pt}
\end{minipage}
\\\\
We will prove that they are derivable from $\dzx^{\textnormal{K,P,E}}_{\frac{\pi}{4}}$.

Some of the rules are already present in $\dzx^{\textnormal{K,P,E}}_{\frac{\pi}{4}}$: \so, \st, \e, \bo, \bt, \h. Thanks to Proposition \ref{prop:TH-derivable-from-clifford-t-toffoli}, we can use Theorem \ref{thm:provability}. This proves that rules (C) and (SUP) are derivable from $\dzx^{\textnormal{K,P,E}}_{\frac{\pi}{4}}$ (Corollaries \ref{cor:axiom-C} and \ref{cor:supplementarity}), as well as (K'). There only remain (EU) and (BW'). \bw appears in $\dzx^{\textnormal{K,P,E}}_{\frac{\pi}{4}}$, but in its ``triangle-form''. We need to prove that the triangle can be decomposed in the diagram of the \frag4 used in \cite{JPV}.\\
The Euler decomposition of the Hadamard gate is derivable:
\def\fig{euler-decomp-proof}
\begin{align*}
\dzx^{\textnormal{K,P,E}}_{\frac{\pi}{4}}\vdash\input{./figures/\fig/\fig_00.tikz}
\eq{\ref{cor:control-alpha-triangles}}\input{./figures/\fig/\fig_01.tikz}
\eq{\ref{lem:parallel-triangle-hadamard}}\input{./figures/\fig/\fig_02.tikz}
\eq{\bt\\\ref{lem:k1}}\input{./figures/\fig/\fig_03.tikz}\\
\eq{\h\\\so}\input{./figures/\fig/\fig_04.tikz}
\eq{\ref{lem:parallel-triangle-hadamard}}\input{./figures/\fig/\fig_05.tikz}
\eq{\ref{cor:control-alpha-triangles}}\input{./figures/\fig/\fig_06.tikz}
\eq{\bo\\\ref{lem:bicolor-0-pi}\\\so\\\st}\input{./figures/\fig/\fig_07.tikz}
\end{align*}
We now have proven all the axioms that make the \frag2 complete. We can use it in the following (denoted $\scomp$).
The decomposition of the triangle node is derivable:
\def\fig{triangle-decomposition-proof}
\begin{align*}
\dzx^{\textnormal{K,P,E}}_{\frac{\pi}{4}}\vdash\input{./figures/\fig/\fig_00.tikz}
\eq{\st\\\so\\\ref{lem:not-triangle-is-symmetrical}}\input{./figures/\fig/\fig_01.tikz}
\eq{\scomp}\input{./figures/\fig/\fig_02.tikz}
\eq{\ref{lem:transistor-ket-1}}\input{./figures/\fig/\fig_03.tikz}
\eq{\ref{lem:triangle-hadamard-2}}\input{./figures/\fig/\fig_04.tikz}
\eq{\text{(EU)}}\input{./figures/\fig/\fig_05.tikz}\\
\eq{\bt}\input{./figures/\fig/\fig_06.tikz}
\eq{\tcx}\input{./figures/\fig/\fig_07.tikz}
\eq{\ref{lem:hopf}\\\so}\input{./figures/\fig/\fig_08.tikz}
\eq{\ref{lem:cnot-on-upside-down-triangle-fork}}\input{./figures/\fig/\fig_09.tikz}\\
\eq{\ref{cor:control-alpha-triangles}}\input{./figures/\fig/\fig_10.tikz}
\eq{\ref{lem:k1}\\\so\\\kt}\input{./figures/\fig/\fig_11.tikz}
\eq{\h}\input{./figures/\fig/\fig_12.tikz}
\eq{\scomp}\input{./figures/\fig/\fig_13.tikz}
\eq{\scomp}\input{./figures/\fig/\fig_14.tikz}\\
\eq{\h\\\scomp}\input{./figures/\fig/\fig_15.tikz}
\eq{\ref{cor:axiom-C2}}\input{./figures/\fig/\fig_16.tikz}
\eq{\so\\\h\\\ref{lem:hopf}}\input{./figures/\fig/\fig_17.tikz}
\eq{\kt}\input{./figures/\fig/\fig_18.tikz}
\end{align*}
The rule (BW') is now easily derivable from the decomposition of the triangle and the rule \bw.
\end{proof}

\begin{proof}[Theorem \ref{thm:general-complete}]
The set of rules in \cite{JPV-universal} is complete in general. $\dzx^{\textnormal{A,K,P,E}}$ already proves all of them except one: 
\InputIfFileExists{axiom-general.tikz}{}{\input{./figures/axiom-general.tikz}}
.\\\\
By Theorem \ref{thm:provability-4}, $\dzx^{\textnormal{A,K,P,E}}$ proves:
\begin{multicols}{2}
\begin{equation}
\label{eq:C2-distributed-through-W-2}
\def\fig{C-half-distributed-through-W}
\input{./figures/\fig/\fig_03.tikz}\eq{}\input{./figures/\fig/\fig_00.tikz}
\end{equation}
\begin{equation}
\label{eq:gen-supp-pi-4}

\InputIfFileExists{gen-supp-pi_4.tikz}{}{\input{./figures/gen-supp-pi_4.tikz}}

\end{equation}
\begin{equation}
\label{eq:gn-through-W}

\InputIfFileExists{gn-through-W.tikz}{}{\input{./figures/gn-through-W.tikz}}

\end{equation}
\begin{equation}
\def\fig{cos-alpha}
\label{eq:cos-alpha}
\input{./figures/\fig/\fig_00.tikz}\eq{}\input{./figures/\fig/\fig_05.tikz}
\end{equation}
\end{multicols}
Now:
\def\fig{axiom-general-proof}
\begin{align*}
\input{./figures/\fig/\fig_00.tikz}
\eq{\bt}\input{./figures/\fig/\fig_01.tikz}
\eq{\bt\\(\ref{eq:gen-supp-pi-4})}\input{./figures/\fig/\fig_02.tikz}
\eq{}\input{./figures/\fig/\fig_03.tikz}\\
\eq{(\ref{eq:cos-alpha})}\input{./figures/\fig/\fig_04.tikz}
\eq{(\ref{eq:gn-through-W})\\\ref{lem:looped-triangle}\\\ref{eq:C2-distributed-through-W-2}}\input{./figures/\fig/\fig_05.tikz}
\eq{(\ref{eq:gn-through-W})\\\tw}\input{./figures/\fig/\fig_06.tikz}
\eq{(\ref{eq:gn-through-W})\\\ref{lem:looped-triangle}}\input{./figures/\fig/\fig_07.tikz}\\
\eq{\textnormal{(A)}}\input{./figures/\fig/\fig_08.tikz}
\eq{\ref{lem:looped-triangle}\\(\ref{eq:gn-through-W})}\input{./figures/\fig/\fig_09.tikz}
\eq{(\ref{eq:cos-alpha})}\input{./figures/\fig/\fig_10.tikz}\\
\eq{}\input{./figures/\fig/\fig_11.tikz}
\eq{(\ref{eq:gen-supp-pi-4})}\input{./figures/\fig/\fig_12.tikz}
\end{align*}
where:
\begin{align*}
2e^{i\theta_3}\cos(\gamma)&=2e^{i(\gamma_1+\frac{\gamma_0}{2})}\left(e^{i\frac{\gamma_0}{2}}+e^{-i\frac{\gamma_0}{2}}\right)=e^{i\gamma_1}\left(1+e^{i\gamma_0}\right)\\
&\!\underset{\textnormal{(A)}}{=}\frac{e^{i(\beta+\theta_2)}}{2}\left(1+e^{-2i\beta}\left(1+e^{i(\alpha+\beta+\theta_1-\theta_2)}\left(1+e^{-2i\alpha}\right)\right)\right)\\
&=\frac{e^{i(\beta+\theta_2)}}{2}\left(1+e^{-2i\beta}\left(1+2e^{i(\beta+\theta_1-\theta_2)}\cos(\alpha)\right)\right)\\
&=\frac{1}{2}\left(e^{i\theta_2}\left(e^{i\beta}+e^{-i\beta}\right)+2e^{i\theta_1}\cos(\alpha)\right)=e^{i\theta_1}\cos(\alpha)+e^{i\theta_2}\cos(\beta)\\
\end{align*}
\end{proof}

\end{document}

%% file: figures/Hadamard-small.tikz
\begin{tikzpicture}
	\begin{pgfonlayer}{nodelayer}
		\node [style=H box] (0) at (0, 0) {};
		\node [style=none] (1) at (0, 0.25) {};
		\node [style=none] (2) at (0, -0.25) {};
	\end{pgfonlayer}
	\begin{pgfonlayer}{edgelayer}
		\draw (1.center) to (2.center);
	\end{pgfonlayer}
\end{tikzpicture}

%% file: figures/Hadamard.tikz
\begin{tikzpicture}
	\begin{pgfonlayer}{nodelayer}
		\node [style={H box}] (0) at (0, 0) {};
		\node [style=none] (1) at (0, 0.5) {};
		\node [style=none] (2) at (0, -0.5) {};
	\end{pgfonlayer}
	\begin{pgfonlayer}{edgelayer}
		\draw (2.center) to (1.center);
	\end{pgfonlayer}
\end{tikzpicture}

%% file: figures/single-line.tikz
\begin{tikzpicture}
	\begin{pgfonlayer}{nodelayer}
		\node [style=none] (0) at (0, 0.2499999) {};
		\node [style=none] (1) at (0, -0.2499999) {};
	\end{pgfonlayer}
	\begin{pgfonlayer}{edgelayer}
		\draw (0.center) to (1.center);
	\end{pgfonlayer}
\end{tikzpicture}

%% file: figures/cup.tikz
\begin{tikzpicture}
	\begin{pgfonlayer}{nodelayer}
		\node [style=none] (0) at (-0.2500001, 0.2500001) {};
		\node [style=none] (1) at (0.2500001, 0.2500001) {};
	\end{pgfonlayer}
	\begin{pgfonlayer}{edgelayer}
		\draw [bend right=90, looseness=1.75] (0.center) to (1.center);
	\end{pgfonlayer}
\end{tikzpicture}

%% file: figures/caps.tikz
\begin{tikzpicture}
	\begin{pgfonlayer}{nodelayer}
		\node [style=none] (0) at (-0.2500001, -0) {};
		\node [style=none] (1) at (0.2500001, -0) {};
	\end{pgfonlayer}
	\begin{pgfonlayer}{edgelayer}
		\draw [bend left=90, looseness=1.75] (0.center) to (1.center);
	\end{pgfonlayer}
\end{tikzpicture}

%% file: figures/triangle.tikz
\begin{tikzpicture}
	\begin{pgfonlayer}{nodelayer}
		\node [style=none] (0) at (0, 0.5) {};
		\node [style=none] (1) at (0, -0.5) {};
		\node [style=ug] (2) at (0, -0) {};
	\end{pgfonlayer}
	\begin{pgfonlayer}{edgelayer}
		\draw [style=none] (0.center) to (1.center);
	\end{pgfonlayer}
\end{tikzpicture}

%% file: figures/Z-1-1.tikz
\begin{tikzpicture}
	\begin{pgfonlayer}{nodelayer}
		\node [style={white dot}] (0) at (0, -0) {};
		\node [style=none] (1) at (0, 0.5) {};
		\node [style=none] (2) at (0, -0.5) {};
	\end{pgfonlayer}
	\begin{pgfonlayer}{edgelayer}
		\draw (1.center) to (2.center);
	\end{pgfonlayer}
\end{tikzpicture}

%% file: figures/W-1-1.tikz
\begin{tikzpicture}
	\begin{pgfonlayer}{nodelayer}
		\node [style=dot] (0) at (0, -0) {};
		\node [style=none] (1) at (0, 0.5) {};
		\node [style=none] (2) at (0, -0.5) {};
	\end{pgfonlayer}
	\begin{pgfonlayer}{edgelayer}
		\draw (1.center) to (2.center);
	\end{pgfonlayer}
\end{tikzpicture}

%% file: figures/W-1-2.tikz
\begin{tikzpicture}
	\begin{pgfonlayer}{nodelayer}
		\node [style=dot] (0) at (0, 0) {};
		\node [style=none] (1) at (0, 0.5000001) {};
		\node [style=none] (2) at (-0.25, -0.5) {};
		\node [style=none] (3) at (0.25, -0.5) {};
	\end{pgfonlayer}
	\begin{pgfonlayer}{edgelayer}
		\draw (0) to (2.center);
		\draw (3.center) to (0);
		\draw (0) to (1.center);
	\end{pgfonlayer}
\end{tikzpicture}

%% file: figures/additional-ZW-rule-zero.tikz
\begin{tikzpicture}
	\begin{pgfonlayer}{nodelayer}
		\node [style=st] (0) at (-1.5, -0) {};
		\node [style=none] (1) at (-0.5000001, -0) {=};
		\node [style=none] (2) at (-0.5, -0.25) {$z$};
		\node [style=dot] (3) at (-1, -0) {};
		\node [style=dot] (4) at (0, -0) {};
	\end{pgfonlayer}
\end{tikzpicture}

%% file: figures/1_sqrt-2-ZX.tikz
\begin{tikzpicture}
	\begin{pgfonlayer}{nodelayer}
		\node [style=rn] (0) at (0, -0.25) {};
		\node [style=gn] (1) at (0, 0.25) {};
	\end{pgfonlayer}
	\begin{pgfonlayer}{edgelayer}
		\draw [bend left=45, looseness=1.00] (0) to (1);
		\draw [bend right=45, looseness=1.00] (0) to (1);
		\draw (0) to (1);
	\end{pgfonlayer}
\end{tikzpicture}

%% file: figures/gn-alpha-0-1.tikz
\begin{tikzpicture}
	\begin{pgfonlayer}{nodelayer}
		\node [style=gn] (0) at (0, 0.125) {$\alpha$};
		\node [style=none] (1) at (0, -0.25) {};
	\end{pgfonlayer}
	\begin{pgfonlayer}{edgelayer}
		\draw [style=none] (0) to (1.center);
	\end{pgfonlayer}
\end{tikzpicture}